\documentclass[%
 reprint,
 amsmath,amssymb,
 aps,
]{revtex4-1}
\usepackage{bm}% bold math
\usepackage{hyperref}% add hypertext capabilities
\usepackage{dblfloatfix}
\usepackage{float}
\usepackage[utf8]{inputenc}
\usepackage{braket}
\usepackage{url}
\usepackage{amsfonts}
\usepackage{bbold}
\usepackage{graphicx}
\usepackage{caption}
\usepackage{subcaption}
\usepackage{appendix}
\usepackage{comment}
\usepackage{float}
\usepackage{amsmath}
\usepackage{tabularx}
\graphicspath{ {images/} }

\begin{document}

\setlength{\abovedisplayskip}{5pt} \setlength{\abovedisplayshortskip}{0pt}

\title{Measurement induced quantum walks}
\author{A. Didi and E. Barkai}
 
\affiliation{%
 Department of Physics, Institute of Nanotechnology and Advanced Materials, Bar-Ilan University, Ramat-Gan 5290002, Israel\\
}%

\date{\today}
\begin{abstract}
We investigate a tight binding quantum walk on a graph.
Repeated stroboscopic measurements of the position of the particle yield a measured "trajectory", and a combination of classical and quantum mechanical properties for the walk are observed.
We explore the effects of the measurements on the spreading of the packet on a one dimensional line, showing that except for the Zeno limit, the system converges to Gaussian statistics similarly to a classical random walk.
A large deviation analysis and an Edgeworth expansion yield quantum corrections to this normal behavior.
We then explore the first passage time to a target state using a generating function method, yielding properties like the quantization of the mean first return time.
In particular, we study the effects of certain sampling rates which cause remarkable change in the behavior in the system, like divergence of the mean detection time in finite systems and a decomposition of the phase space into mutually exclusive regions, an effect that mimics ergodicity breaking, whose origin here is the destructive interference in quantum mechanics. 
For a quantum walk on a line we show that in our system the first detection probability decays classically like $(\text{time})^{-3/2}$, this is dramatically different compared to local measurements which yield a decay rate of $(\text{time})^{-3}$, indicating that the exponents of the first passage time depends on the type of measurements used.
\end{abstract} 

\pacs{Valid PACS appear here}
                             
\maketitle

\section{Introduction}
Consider a non-biased random walk in dimension one, for example a walk with Gaussian increments, or random walk on a lattice with equal probability to jump left or right.
In the long time limit the density of particles all starting at the origin will converge to a Gaussian distribution.
On the other hand, for a tight binding quantum walk on a line with nearest neighbor jumps, it is well known that the PDF, given by the square modulus of the wave function, will propagate ballistically in the absence of measurement.
The question we seek to answer is, what will happen in the case where we measure the position of the particle stroboscopically, as explained in the abstract?
These measurements yield a path that the particle took, given by the sites that it was detected at.
From this path we may construct in principle the probability of finding the particle on lattice point x after n measurements.
Will the repeated collapse of the wave function cause the statistics of the system to display features typical of a classical random walk, or will the process remain a purely quantum one in spite of this, or some combination of the two?
In our work we found that the measurements induce a Gaussian behavior, similar to what one would expect of a classical random walk, however a closer look at the problem using large-deviation theory and an Edgeworth expansion reveal the quantum nature of the process. 
Notwithstanding these effects, the convergence to the Gaussian behavior is typically fast, with the exception of the Zeno limit $\tau \to 0$, where the non-classical effects are better pronounced.

\begin{figure}[h]
  \centering
  \includegraphics[width=0.48\textwidth]{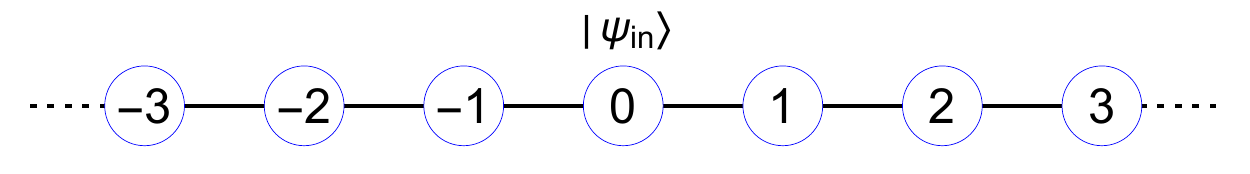}
  \caption{
  A segment of the infinite line.
  For a particle starting at the origin, we allow it to unitarily evolve for $\tau$ time, and then we measure it's position.
  The measurement localizes it's wave function to some single site on the lattice, which we record as being the location of the particle at time $\tau$.
  After this we allow the particle to freely evolve for $\tau$ time once more before we measure it's position again.
  By repeating this process many times, we generate a "path" that the particle took.
  For the lattice shown here, an example of a path would be:
  $\{ 0, 2, 1, 0, -3, ... \}$
  }
  \label{fig:inflinegraphix}
\end{figure}

To more precisely define our model, we examine a closed quantum system which is prepared in some definite initial state and evolves unitarily for a predetermined time $\tau$, at which point we measure it's position.
The free evolution between measurements is given by unitary dynamics determined by the Hamiltonian of the system, typically described by the adjacency matrix of a graph.
After measurement is complete we log the location that we found the particle in and then allow it to freely evolve for $\tau$ time once more, before repeating the process, such as in the case presented in Fig \ref{fig:inflinegraphix}.
Our goals in this work are twofold, to better understand the statistical properties of such a process were it to repeat indefinitely, and to examine the first detection statistics when we are interested in the amount of time it will take for the system to reach a certain state under this measurement protocol.
An analogous classical example \cite{SidRedner} would be to record the position of an animal randomly wandering in the wilderness at a constant frequency $1/\tau$.
By doing this many times, we can define a path that the animal took as the list of positions we recorded, with this we can also ask questions about the statistics of how long it would take the animal to travel between two different locations, or to return to it's starting place.
Where this analogy breaks down however, is that while for an animal the act of recording its position and the frequency of recording wouldn't change it's path, in our case it is the driving force of the random walk, as we'll later show.

The rest of the paper is organized as follows:
In Secs. \ref{section:model}, \ref{sec:probabilityvector} we define our model with regard to both the measurement induced quantum random walk without termination and the first detection problem.
In Sec. \ref{sec:genfuncsec} we describe the mathematical framework we'll be using and in Secs. \ref{sec:momentresults}, \ref{sec:slowrelatxionsection} we use it to derive several properties of this type of random walk.
In Sec. \ref{sec:examples} we apply the aforementioned results and mathematical tools to study a measurement induced quantum random walk finite graph in detail and in Sec. \ref{sec:infiniteline} we do the same for an infinite 1d lattice.
We follow this up in Sec. \ref{section:comparison} with a comparison of our measurement scheme to a local measurement model, as such models have seen extensive research in papers by Dhar et al. \cite{Dhar1, Dhar2}, Krovi and Brun \cite{krovi1, krovi2}, and others.
In particular, we'll focus on directly comparing our results to those found for the scheme considered in \cite{PhysRevE.95.032141, PhysRevResearch.2.033113, PhysRevResearch.1.033086}.
We close the paper with discussions and a summary in Sec. \ref{sec:summarysec}.
Detailed calculations and an additional example are presented in the appendices.

\section{Model}
\label{section:model}
Our model can broadly be divided into two primary elements.
The first are the quantum dynamics described by the Schr\"odinger equation.
The second being the measurement protocol which yields the well defined meaning for the position of the particle.
Beginning with a description of the first part, we consider a single particle whose time evolution is described by a time independent Hermitian Hamiltonian $H$ according to the Schr\"odinger equation $H\ket{\psi} = i\hbar\frac{d}{dt}\ket{\psi}$ where we set $\hbar=1$.
The initial wave function is denoted $\ket{\psi_{in}}$.
The $H$ is represented here with a graph where nodes describe states and edges describe the hopping amplitudes between these states.
Our main focus considers a $H$ which is described by some adjacency matrix, though the theory presented below is more general.

\begin{figure}[h]
     \centering
         \centering
         \includegraphics[width=0.3\textwidth]{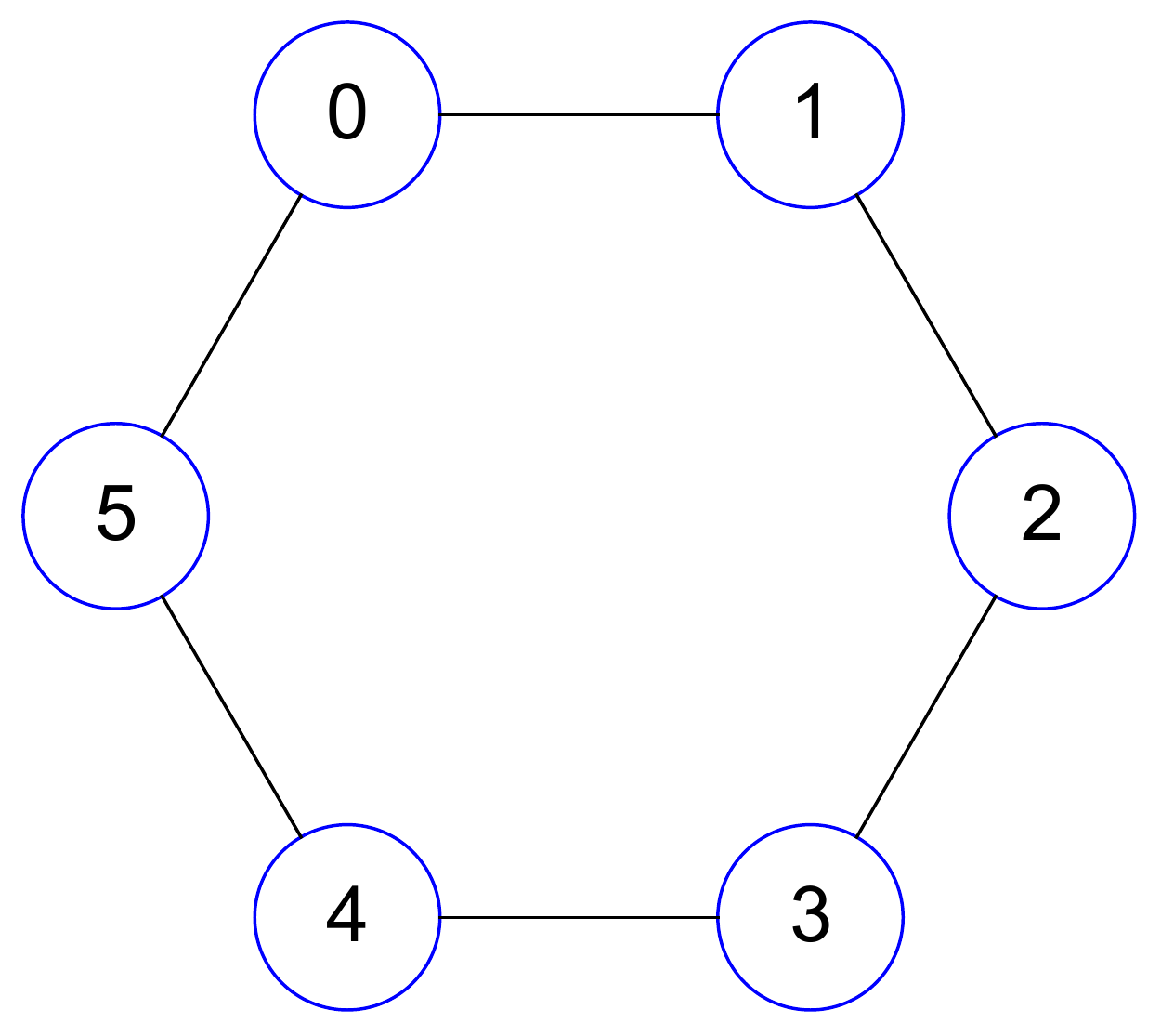}
         \caption{Schematic model of a benzene ring. The edges represent jumps between nearest neighbours, all with the same amplitude. In Sec \ref{sec:examples} we examine the properties of a measurement induced quantum walk on this graph, as well as the first detection problem.}
         \label{fig:Hexagon}
\end{figure}

The Hilbert space we consider is discrete and is spanned by $X$ which is the set of eigenvectors of the operator $\hat{X}$.
$\hat{X}$ can be any arbitrary Hermitian operator as long as all of its eigenvalues are non-degenerate.
In practice, in this paper we'll assume that it is the position operator for the sake of simplicity.
The initial wave function $\ket{\psi_{in}}$ is localized to a single site in this space $\ket{\psi_{in}} \in X$.
For example, we consider the tight binding infinite line lattice Hamiltonian
\begin{equation}
\label{eq:inflineH1stApperance}
    H=-\gamma \sum_{x=-\infty}^{\infty} (\ket{x}\bra{x+1} + \ket{x+1}\bra{x})
\end{equation}
where $\ket{\psi_{in}} =\ket{0}$.
This Hamiltonian describes hops between nearest neighbours on the infinite line with an amplitude $\gamma$ for the jumps, and hence for an initial condition set at the origin these dynamics are sometimes called a tight binding quantum walk \cite{blumenohysicsreport}.
In this example the space $X$ is defined with the vertices on the lattice $\ket{x}$ and as usual $\hat{X}\ket{x} = x \ket{x}$.
A schematic representation of this Hamiltonian is given in Fig. \ref{fig:inflinegraphix}.
Another example is presented in Fig. \ref{fig:Hexagon} where we have a finite ring.
It is well known that the statistical properties of first detection time for finite and infinite systems dramatically differ \cite{SidRedner, PhysRevE.95.032141, PhysRevLett.120.040502} , so we will later use these two models as examples in sections \ref{sec:examples} and \ref{sec:infiniteline}.

\begin{figure}[h]
  \centering
  \includegraphics[width=0.48\textwidth]{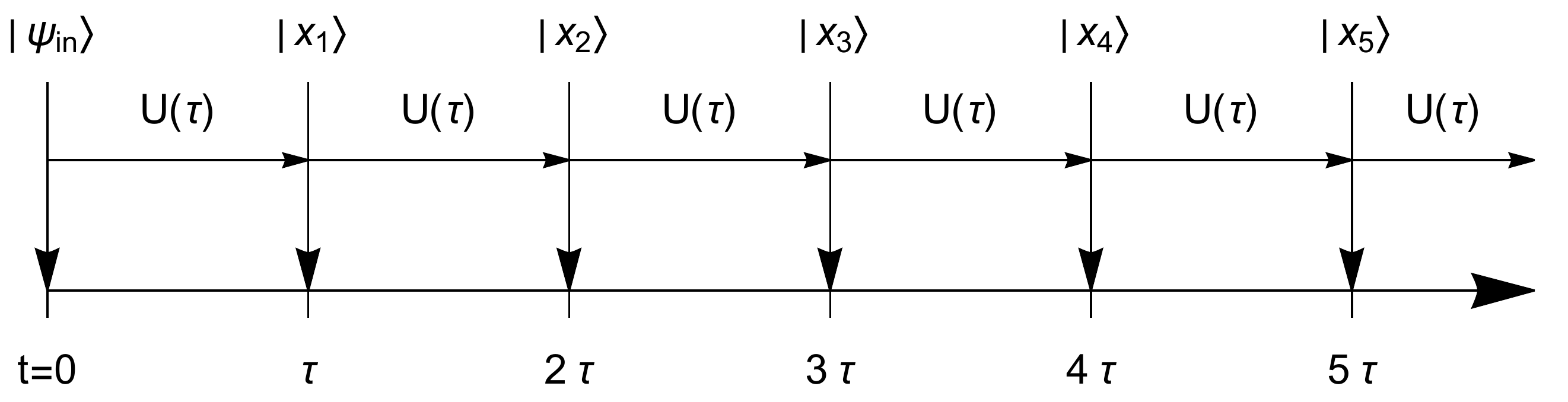}
  \caption{A measurement induced quantum walk which begins at $\ket{\psi(t=0)} = \ket{\psi_{in}}$. At every step of the walk the particle is detected at some point on the graph $\ket{x_n}$ and between each detection the wave function evolves according to the Schr\"odinger equation with the standard unitary time evolution operator $U(\tau) = e^{-i \tau H}$.
  This process can, in principle, continue indefinitely.}
  \label{fig:MeasuerementVisualizationFigure2}
\end{figure}

In order to establish the position of the particle at every step of the random walk, we start by defining the measurement protocol as follows:
Position measurements are performed at discrete times $t=\tau, 2\tau,... N\tau$ using the position operator $\hat{X}$.
Between each measurement event the dynamics are unitary as described above. 
At every measurement we obtain the exact position of the particle causing the wave function to collapse and be localized to that point on the graph.
Of course, the basic postulates of quantum mechanics mean that the outcome of the measurement is random.
Over the course of many measurements this process produces a list of the locations the particle was detected at.
As the measurements combined with the unitary time evolution given by the Schr\"odinger equation act as the generators of this random walk, we shall henceforth refer to this process as a measurement induced quantum walk.
As an example of a particular instance of this measurement induced quantum walk, on the infinite line Hamiltonian in Eq. (\ref{eq:inflineH1stApperance}) the Eigenvalues of the position operator are the integers $X = \mathbb{Z}$ so for a walk whose initial state is the origin ($\ket{\psi_{in}} = \ket{0}$) one possible sequence of measurement outcomes is $\{ x_1 = 1, x_2 = 3, x_3 = 1, x_4 = -1, x_5 = ... \}$.
For a more general measurement induced quantum walk, the unitary time evolution followed by measurement induced wave function collapsed repeats indefinitely, as is shown in Fig \ref{fig:MeasuerementVisualizationFigure2}.

To be more precise with regard to the exact mechanics of the measurement induced quantum walk, consider the first measurement which occurs at $t=\tau$.
Immediately before this time at $\tau^-=\tau-\epsilon (\epsilon \to 0)$ the wave function is given by:
\begin{equation}
    \ket{\psi(\tau^-)}=e^{-i \tau^- H} \ket{\psi_{in}}
    .
\end{equation}
According to the standard interpretation of measurement in quantum mechanics, the detection probability at $t=\tau$ at every graph site $x$ is given by:
\begin{equation}
\label{eq:p1eq1}
    P_1=|\braket{x|\psi(\tau^-)}|^2
    .
\end{equation}
After measurement the particle will be localized to some site which we'll label as $x_1$.
Meaning that at time $\tau^+=\tau+\epsilon$ the wave function is given by:
\begin{equation}
    \ket{\psi(\tau^+)}=\ket{x_1}
    .
\end{equation}
We then allow the system to freely evolve in time until $2\tau^-=2\tau-\epsilon$ when it is $\ket{\psi(2\tau^-)}=e^{-i \tau^- H} \ket{x_1}$.
We then measure the particle again, and the probability of detection for each site $x$ is given by:
\begin{equation}
    P_2=|\braket{x|\psi(2\tau^-)}|^2
    .
\end{equation}
The particle is then localized to a new state which we'll label as $x_2$.
We continue to evolve the wave function and measure its position in this manner, obtaining a list of its positions in the process $\{ x_1, x_2, x_3 ... \}$ which we treat as being the positions that the particle traveled through over the course of its random walk.

\section{Dynamics with Measurements}
\label{sec:probabilityvector}
The probability of detecting the particle at the state $x$ at time $\tau n$ is given by the sum of the probabilities of all paths which begin at the origin $\psi_{in}$ and reach $x$ after $n$ steps.
In order to find these probabilities we define a vector which we'll call the probability vector, which will contain the probabilities of the possible outcomes of the measurements at times which are integer multiples of $\tau$ plus an infinitesimally small positive $\epsilon$.
Since this vector is just the main diagonal of the density matrix describing the system, we'll denote it as $\ket{\rho(\tau n)}$.
It is a non negative vector whose dimension is the same as that of the Hilbert space, but it's certainly not part of the Hilbert space.
Since the total probability that the particle will be found somewhere on the graph is unity, the sum of the elements of the probability vector equals one.

At time $t=0$ we define the probability vector to equal $\ket{\psi_{in}}$, we assume that it is an eigenstate of $\hat{X}$.
In the context of tight binding walks on a graph, this means that the system is initially localised to a node of the graph which we call $\ket{\psi_{in}}$.
We address the case where it is not initially localized in appendix \ref{appendix:nonlocal}.
For the time evolution of $\ket{\rho(\tau n)}$ we define the operator $G$ such that it will contain the probabilities for the particle to jump from any position in the system to any other position:
\begin{equation}
\label{eq:Gdefinition}
    G = \sum_{x, x' \in X} |\braket{x'|e^{-i \tau H}|x}|^2 \ket{x'} \bra{x}
    .
\end{equation}
$G$ is a stochastic matrix, meaning that it is used to describe the transitions of a Markov chain.
All of $G$'s eigenvalues are real and have an absolute value less than or equal to one, as is shown in appendix \ref{appendix:eigenG}.
Evolving the system in time using this matrix, the state of the system at time $t = \tau n$ is described by the probability vector:
\begin{equation}
\label{eq:probvecwithoutabsorbingboundaryconditions}
    \ket{\rho (\tau n)} = G^n \ket{\psi_{in}}
    .
\end{equation}
This is a kind of discrete time Master equation which takes into consideration both the unitary time evolution and the periodic measurements that collapse the wave function.

\begin{figure}[h]
  \centering
  \includegraphics[width=0.48\textwidth]{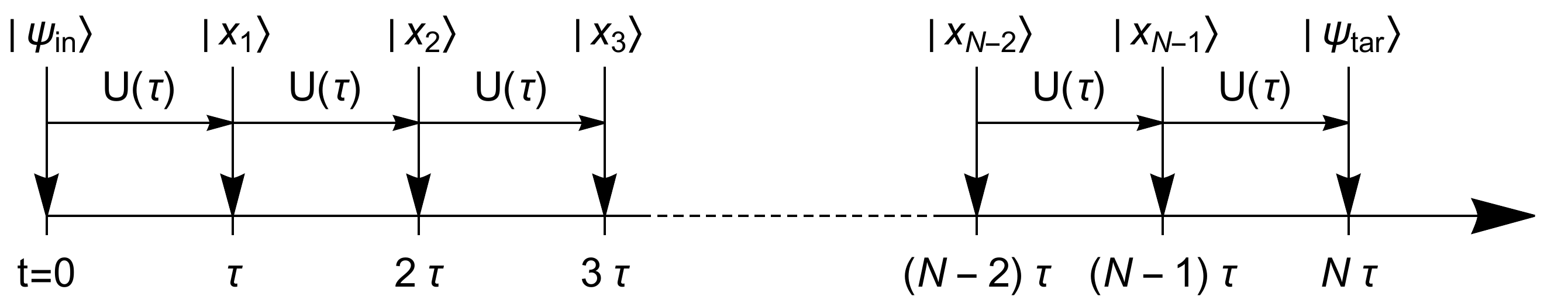}
  \caption{A measurement induced quantum walk which begins at $\ket{\psi(t=0)} = \ket{\psi_{in}}$ and ends after time $t=N \tau$ when the particle is detected at $\ket{\psi_{tar}}$. The underlying physical process is the same as the one that was described in Fig. \ref{fig:MeasuerementVisualizationFigure2} but in this process we stop the random walk once it's detected at the target state $\ket{\psi_{tar}}$.}
  \label{fig:MeasuerementVisualizationFigure}
\end{figure}

In addition to the measurement induced quantum walk itself, another topic of interest we study in this paper is the first detection problem, wherein rather than simply allow the system to continue evolving in time indefinitely we define a target state $\ket{\psi_{tar}}$ and we stop the random walk once the particle is detected in this state.
This state can either be the same as the origin, in which case the random walk is referred to as a return problem, or it can be any other state on the graph in which case we refer to it as a transition problem.
The first hitting time is $N \tau$, and it is a random variable whose statistical properties depend on the Hamiltonian and the particular choice of sampling interval $\tau$.
The modified process for this random walk is described in Fig. \ref{fig:MeasuerementVisualizationFigure}.
To account for the fact that in this version of the measurement induced quantum walk the experiment stops when the particle is detected at $\psi_{tar}$ we remove the probability that the particle was found at $\psi_{tar}$ after every $\tau$ since the continuation of the experiment necessarily means it was not found there.
This effectively removes any paths that crossed through $\psi_{tar}$ from our ensemble since those paths would've resulted in the termination of the process.
Projecting the resulting probability vector unto the target state gives us the following equation for the probability of \emph{first} detecting the particle at the target site at time $t=\tau n$, which we denote as $F_n$:
\begin{equation}
\label{eq:Fn1}
    F_n = \braket{\psi_{tar}| (G(1-D))^{n-1}G | \psi_{in}}
    .
\end{equation}
Where $D = \ket{\psi_{tar}}\bra{\psi_{tar}}$ is a projecting operator.
For example, for $n=1$, we obtain $F_1=\braket{\psi_{tar}|G|\psi_{in}}$.
For $n=2$, we have $F_2=\braket{\psi_{tar}|G(1-\ket{\psi_{tar}}\bra{\psi_{tar}})G|\psi_{in}}$ and so on.
The full derivation of Eq. (\ref{eq:Fn1}) is presented in appendix \ref{appendix:Fn}.
We will now further interpret Eq. (\ref{eq:Fn1}) using a renewal equation approach.

\section{Generating function}
\label{sec:genfuncsec}
In appendix \ref{appendix:equationEquivelance} we show that Eq. (\ref{eq:Fn1}) is equivalent to:
\begin{equation}
    \label{eq:Fn2}
    F_n = \braket{\psi_{tar}|G^n|\psi_{in}} - \sum_{j=1}^{n-1} F_{j}\braket{\psi_{tar}|G^{n-j}|\psi_{tar}}
    .
\end{equation}
This is a renewal equation typically found for these types of problems \cite{SidRedner, PhysRevE.95.032141}.
One may obtain a simple physical interpretation of this equation by moving the sum to the left hand side, which gives us:
\begin{equation}
    \label{eq:Fn2otherform}
    \sum_{j=1}^{n} F_{j}\braket{\psi_{tar}|G^{n-j}|\psi_{tar}} = \braket{\psi_{tar}|G^n|\psi_{in}}
    .
\end{equation}
In doing so, we can see that the probability of arriving at the target state after $n$ steps $\bra{\psi_{tar}} G^n \ket{\psi_{in}}$ (not necessarily for the first time) is the same as the probability of arriving there for the first time at some earlier point in time and then looping back to return there.

In order to analyse $F_n$ we note that Eq. (\ref{eq:Fn2}) has a convolution structure hence we use the Z transform, or discrete Laplace transform, which is by definition \cite{doi:10.1021/ar050028l}:
\begin{equation}
\label{eq:ogGenFuncDefinition}
    \widetilde{F(z)} = \sum_{n=1}^{\infty} F_{n} z^{n}
    .
\end{equation}
Multiplying Eq. (\ref{eq:Fn2}) by $z^n$ and summing over $n$ we get:
\begin{equation}
    \widetilde{F(z)} = \sum_{n=1}^{\infty} \braket{\psi_{tar}|G^n|\psi_{in}} - \sum_{n=1}^{\infty} \sum_{j=1}^{n-1} F_{j}\braket{\psi_{tar}|G^{n-j}|\psi_{tar}}
    .
\end{equation}
The second term in the right hand side of the equation is a discrete convolution, so after evaluating the sums over $n$ and some rearrangement we get:
\begin{equation}
    \label{GeneratingFunction}
    \widetilde{F(z)} = \frac{\braket{\psi_{tar}|\widetilde{G(z)}|\psi_{in}}}{1 + \braket{\psi_{tar}|\widetilde{G(z)}|\psi_{tar}}}
\end{equation}
Where $\widetilde{G(z)} = \sum_{n=1}^{\infty} z^{n}G^{n} = zG/(1-zG)$.
By rearranging the generating function, we are also able to present it in a manner which better relates it to the underlying probability vector:
\begin{equation}
    \label{eq:GeneratingFunctionProbVecVersion}
    \widetilde{F(z)} = \frac{\sum_{n=1}^{\infty} z^n \braket{\psi_{tar}|\rho_{in}(\tau n)}}{\sum_{n=0}^{\infty} z^n \braket{\psi_{tar}|\rho_{tar}(\tau n)}}
\end{equation}
Where $\ket{\rho_{in}(0)} = \ket{\psi_{in}}$ and $\ket{\rho_{tar}(0)} = \ket{\psi_{tar}}$.
In the return problem where $\ket{\psi_{in}} = \ket{\psi_{tar}}$, this expression can be further simplified to just:
\begin{equation}
    \label{eq:GeneratingFunctionProbVecVersion2RETURN}
    \widetilde{F(z)} = 1-\frac{1}{\sum_{n=0}^{\infty} z^n \braket{\psi_{in}|\rho(\tau n)}}
    .
\end{equation}

Using the generating function we are able to calculate the first detection probabilities, the survival probability, expected number of measurement attempts, and the variance in the number of detection attempts.
We formally find that they are given by:
\begin{equation}
\label{eq:F_nEq}
    F_n=\frac{1}{n!} \frac{d^n}{dz^n} \widetilde{F(z)} |_{z=0}=\frac{1}{2 \pi i} \oint _ {C } \frac{\widetilde{F(z)}}{z^{n+1}} dz
\end{equation}
\begin{equation}
    \label{<n^0>}
    P_{det} = \sum_{n=1}^{\infty} F_n = \widetilde{F(1)}
\end{equation}
\begin{equation}
    \label{<n^1>}
    \braket{n} = \sum_{n=1}^{\infty} n F_n = \frac{d}{dz} \widetilde{F(z)} |_{z=1}
\end{equation}
\begin{equation}
    \label{<n^2>}
    \braket{n^2} = \sum_{n=1}^{\infty} n^2 F_n = \frac{d}{dz} \left( z\frac{d}{dz} \widetilde{F(z)} \right)|_{z=1}
\end{equation}
Here $P_{det}$ is the total detection probability after an infinite number of measurement attempts.
If $P_{det}$ is 1 Eq. (\ref{<n^1>}) and (\ref{<n^2>}) are the moments of the first detection event.
Note that like in classical random walks the total detection probability $P_{det}$ can be less than unity \cite{SidRedner, PhysRevE.95.032141}, in which case the first and second moment can no longer be used to calculate the average and variance of the first detection event as detection is not guaranteed.
They can still be used to compute the average and variance conditioned on the event that the particle is detected by dividing them by the total detection probability: $\text{Average} = \braket{n}/P_{det}$ and $\text{Variance} = \braket{n^2}/P_{det} - \braket{n}^2/P_{det}^2$.

So far, the results we've obtained seem to indicate that the measurement induced quantum walk behaves like a regular discrete time classical random walk where other than the fact that the transition probabilities are determined using the Schr\"odinger equation the process is completely classical.
By this we mean that the basic structure of the renewal equation  is classical.
This classical feature is clearly related to the repeated measurements which help us define a discrete path on the graph that the particle took, a feature which is classical.
While the aforementioned classical trait would seem to imply that this process is purely classical, this ignores some interesting effects that arise from the combination of the unitary time evolution of the quantum wave function with the wave function collapse introduced by repeated measurements.
First, $G$ depends of course on $\hbar$ and in that sense it is still describing a quantum mechanical process.
But more profoundly, we find that for certain sampling rates which we label as exceptional the statistics of the system vary drastically compared to classical behavior, and that depending on the initial and target states certain sampling rates minimize the average time until detection whereas for others the average diverges to infinity.
Such features are different compared to classical walks on similar structures, and are related to periodicity, revivals, and destructive interference of the underlying quantum dynamics.
These underlying dynamics embedded in $G$ through the unitary time evolution of the wave function $U(\tau)$ make it fundamentally different from classical transition matrices in spite of the superficial similarities.

\section{Expectation values and exceptional sampling rates} \label{sec:momentresults}
In this section we derive general expressions for $P_{det}$, $\braket{n}$, and $\Delta n^2$ in finite systems.
We also observe interesting effects in these values near exceptional sampling rates, examples of which are later shown in Sec. \ref{sec:examples}.
\emph{
We define these sampling rates to be those that cause 1 to be a degenerate eigenvalue of $G$.
}
We postulate that these exceptional sampling rates satisfy $\Delta E \tau = 2 \pi n$ where $\Delta E$ are the non-negative differences between pairs of the Hamiltonians eigen-energies.
\subsection{Total detection probability}
As previously mentioned in Eq. (\ref{<n^0>}), the total detection probability is acquired by evaluating the generating function at $z=1$, in this section we derive a general expression for this.
Since $G$ is a real Hermitian matrix, we can express the initial and detection sites as super-positions of it's eigenstates.
\begin{equation}
\label{eq:inanddetexpression1}
\begin{split}
    \ket{\psi_{in}} &= \sum_{\lambda} \sum_{k=1}^{g_{\lambda}} \braket{\lambda_k|\psi_{in}}\ket{\lambda_k}
\\
    \ket{\psi_{tar}} &= \sum_{\lambda} \sum_{k=1}^{g_{\lambda}} \braket{\lambda_k|\psi_{tar}}\ket{\lambda_k}
    .
\end{split}
\end{equation}
Where $G\ket{\lambda_k} = \lambda\ket{\lambda_k}$ and $g_{\lambda}$ are the degeneracies of the eigenvalues.
In addition to this general form of writing the initial and target state, we also consider the state:
\begin{equation}
\label{eq:uniformdistributionstate}
    \ket{\phi} = \frac{1}{\sqrt{|X|}}\sum_{x \in X}\ket{x}
    .
\end{equation}
This is an eigenstate of $G$ with eigenvalue one at all sampling rates.
By the prior definition of exceptional sampling rates, for non-exceptional sampling rates $\ket{\phi}$ is the \emph{only} $G$ eigenstate whose eigenvalue is one.
This state serves like a kind of a ground state under the repeated measurements.
In classical processes this would correspond to a high temperature limit (in a Boltzmann sense) since the system is evenly populating all states.
The repeated measurements drive the system to this state \cite{PhysRevA.84.032121}.

Returning to the subject of the derivation of a general formula for $P_{det}$, we plug Eq. (\ref{eq:inanddetexpression1}) into Eq. (\ref{GeneratingFunction}) to obtain:
\begin{equation}
\label{eq:expandedgenfunc}
    \widetilde{F(z)} = \frac{
    \sum_{\lambda} \sum_{k=1}^{g_{\lambda}} \frac{\braket{\psi_{tar}|\lambda_k}\braket{\lambda_k|\psi_{in}} \lambda z}{1-\lambda z} 
    }{
    1 + \sum_{\lambda} \sum_{k=1}^{g_{\lambda}} \frac{|\braket{\lambda_k|\psi_{tar}}|^{2} \lambda z}{1-\lambda z}
    }
    .
\end{equation}
Using Eq. (\ref{<n^0>}) we take the limit $z \to 1$, all of the eigenstates whose eigenvalues are less than one disappear and we are left with:
\begin{equation}
    \label{eq:pdetequation}
    P_{det} = \widetilde{F(1)} = \frac
    {\sum_{k=1}^{g_1} \braket{\psi_{tar}|1_k}\braket{1_k|\psi_{in}}}
    {\sum_{k=1}^{g_1} |\braket{1_k|\psi_{tar}}|^{2}}
    .
\end{equation}
where the summation is only over the set of eigenstates whose eigenvalue is one $G \ket{1_k} = \ket{1_k}$.
This equation is general and valid for all sampling rates.

We can see from this expression that in the return problem, where we set $\braket{\psi_{in}|\psi_{tar}} = 1$, the total detection probability $P_{det}$ is always one in a finite system.
We can also see that for non-exceptional sampling rates, i.e sampling frequencies where 1 is a non-degenerate eigenvalue of $G$, i.e $g_1 = 1$ ,$P_{det}$ is one for the transition problem as well in finite systems, since in that case the only eigenstate which will go into the sum in Eq. (\ref{eq:pdetequation}) is the uniform state given in Eq. (\ref{eq:uniformdistributionstate}).
Thus, for non-exceptional sampling rates the total detection probability is unity just like a regular classical random walk on a finite graph.
This is very different from the case where we only measure locally at the target site where due to destructive interference we may get $P_{det} < 1$ \cite{thiel2019quantum, darkStatesPaper2, darkStatesPaper3}.
We address this subject and the more general comparison between global and local measurements in Sec. \ref{section:comparison}.
\subsection{The mean and variance in the return problem}
The mean $\braket{n}$ for a measurement induced quantum walk on a finite graph is the average number of attempts needed until the particle is first found at the state $\ket{\psi_{tar}}$.
It is equal to the average time from the beginning of the walk until the particle is detected at the target state, divided by the time between measurements $\tau$.
Similarly to the derivation of Eq. (\ref{eq:pdetequation}), we express the initial and detection site as super-positions of $G$'s eigenstates, plug those into Eq. (\ref{<n^1>}) and take the limit as $z \to 1$ to find that in the return problem $\braket{\psi_{in}|\psi_{tar}} = 1$ this average is:
\begin{equation}
\label{eq:avgnfinalform}
    \braket{n} = \frac{1}{\sum_{k=1}^{g_1} |\braket{\psi_{in}|1_k}|^2 }
    .
\end{equation}

For non-exceptional sampling rates $\ket{\phi}$ is the only state in the sum and we can simplify Eq. (\ref{eq:avgnfinalform}) to obtain:
\begin{equation}
\label{eq:avgnfinalform2}
    \braket{n}=|X|
    .
\end{equation}
This means that the mean $\braket{n}$ is quantized and independent of the details of the system besides the dimension of the Hilbert space.
Namely, recall that for any tight binding walk on a graph $X$ is the set of vertexes and $|X|$ is the number of vertices.

In essence this is similar to the Kac formula for the mean, which applies for classical walks and reads $\braket{n} = 1/p_{eq(x)}$ where $p_{eq(x)}$ is the equilibrium state and $x$ is the vertex \cite{B_nichou_2015}.
However, what is remarkable here is that the effective steady state is uniform (like a high temperature limit in classical statistical mechanics), so the corresponding equilibrium measure is $1/|X|$.
Under the condition that $g_1 = 1$, we can interpret the resulting Eq. (\ref{eq:avgnfinalform2}) as the measurements driving the system to a high temperature classical limit \cite{PhysRevA.84.032121}.

One exceptional sampling rate which we can see in all systems is the Zeno limit $(\tau \to 0)$.
In this limit of Eq. (\ref{eq:avgnfinalform}) $G$ becomes the identity matrix which causes the sum in the divisor to evaluate to 1 and we get $\braket{n}=1$.
This is expected, if we start at the target node and immediately measure we record the particle in the first attempt.
What is remarkable is that if $\tau$ is very small but finite we still get $\braket{n} = |X|$.
So close to the Zeno limit we get a drastic jump in $\braket{n}$ when plotted as a function of $\tau$.
This means that the fluctuation of the first detection time $N \tau$ close to this limit are gigantic.
We can also see these fluctuations in that the variance $\Delta n^2$ diverges at this limit as shown in Eq. (\ref{eq:varianceformula}) presented soon.
This non-analytical behavior is also found for other exceptional sampling times, discussed below. 

To find the variance in the number of measurements $\Delta n^2 = \braket{n^2} - \braket{n}^2$ we repeat the aforementioned process for Eq. (\ref{<n^2>}) and find that for non-exceptional sampling rates the variance of the number of measurements in a finite system is given by:
\begin{equation}
    \label{eq:varianceformula}
    \Delta n^2 = |X|^2-|X| + 2 |X|^2\sum_{\lambda \neq 1} \sum_{k=1}^{g_{\lambda}} |\braket{\psi_{in}|\lambda_k}|^2 \frac{\lambda }{1-\lambda}
    .
\end{equation}
We see here that as any of $G$'s eigenvalues $\lambda$ approach one such as in the Zeno limit the variance diverges, which is indicative of the large fluctuations in the first detection probability which are observed near exceptional sampling rates.
Keep in mind that Eq. (\ref{eq:avgnfinalform}) and (\ref{eq:varianceformula}) are only valid in the return problem $\braket{\psi_{in}|\psi_{tar}} = 1$.
In appendix \ref{app:derivationLONG} we present the full derivation of these two equations in detail, as well as a more general form of the equation for $\Delta n^2$ which also works for exceptional sampling rates.

Note that the quantization of the mean return time, Eq. (\ref{eq:avgnfinalform2}) is not unique to the protocol under study.
In Sec. \ref{section:comparison} we discuss other measurement schemes that yield quantization of mean return time.

\subsection{The mean in the transition problem}
We now turn our attention to the transition problem.
Similarly to the previous derivation of Eq. (\ref{eq:avgnfinalform}) we
plug Eq. (\ref{eq:expandedgenfunc}) into Eq. (\ref{<n^1>}).
After simplifying we arrive at the following equation for $\braket{n}$ in the transition problem:
\begin{equation}
    \label{generalBKN}
    \braket{n} = \frac{
    \sum_{k=1}^{g_1} \braket{\psi_{tar}|1_k} \braket{1_k|\psi_{in}} 
    }{
    (\sum_{k=1}^{g_1} |\braket{\psi_{tar}|1_k}|^2)^2
    }
    +
    \frac{\lim_{z \to 1} g(z)}{(\sum_{k=1}^{g_1} |\braket{\psi_{tar}|1_k}|^2)^2}
    .
\end{equation}
The function $g(z)$ is:
\begin{equation}
\begin{split}
    g(z)
    =&\sum_{\lambda} \sum_{k=1}^{g_{\lambda}} \sum_{j=1}^{g_1} \frac{
    \lambda z
    }{
    1-\lambda z
    }
    f(\ket{\lambda_{k}}, \ket{1_{j}})
\\
    f(\ket{\lambda_{k}}, \ket{1_{j}})
    =&\braket{\psi_{tar}|1_{j}} \braket{1_{j}|\psi_{in}} |\braket{\psi_{tar}|\lambda_{k}}|^2\\
    -&\braket{\psi_{tar}|\lambda_{k}} \braket{\lambda_{k}|\psi_{in}} |\braket{\psi_{tar}|1_{j}}|^2
    .
\end{split}
\end{equation}
where the sums in Eq. (\ref{generalBKN}) are all only over eigenstates whose eigenvalue is one and $G\ket{1_j} = \ket{1_j}$.
In $g(z)$ the summation over $\lambda_{k}$ includes all of $G$'s eigenstates whereas the sum over $j$ is over only the eigenvectors whose eigenvalue is one.
For non exceptional sampling rates, the sum over $j$ is just $\ket{\phi}$, and since $\forall x \in X:f(\ket{x}, \ket{x})=0$ Eq. (\ref{generalBKN}) reduces to:
\begin{equation}
\label{NonExeTransitionAvg}
\begin{split}
    \braket{n} &= |X| \huge{\text{[}} 1 + \sum_{\lambda \neq 1} \sum_{k=1}^{g_{\lambda}}
    \\
    & \frac{\lambda}{1-\lambda} (
    |\braket{\psi_{tar}|\lambda_k}|^2 - \braket{\psi_{tar}|\lambda_k} \braket{\lambda_k|\psi_{in}}
    ) \huge{\text{]}}
    .
\end{split}
\end{equation}
If $\lambda \to 1$, $\braket{n}$ can diverge. However, it should be noted that $\braket{n}$ does not necessarily diverge close to exceptional sampling rates since the expression in the inner brackets can equal zero. If it doesn't diverge it often discontinuously jumps to a different finite value. We address this issue in appendix \ref{appendix:n^1 disjump}.
It should also be noted that if $\ket{\psi_{in}} = \ket{\psi_{tar}}$ the sum reduces to zero and we get Eq. (\ref{eq:avgnfinalform2}), since it is just a special case of this equation.

\section{Slow relaxation of the survival probability near exceptional sampling rates}
\label{sec:slowrelatxionsection}
In Sec. \ref{sec:momentresults} we've already shown that in finite systems, as the number of measurements goes to infinity the total detection probability $P_{det}$ converges to one for non-exceptional sampling rates.
In this section we show that this relaxation is considerably slower for sampling rates which are close to exceptional ones.

Starting with Eq. (\ref{eq:Fn1}), we'll define a new operator $G_{tar} = G(\mathbb{1}-\ket{\psi_{tar}}\bra{\psi_{tar}})$ and rewrite the formula with it:
\begin{equation}
    F_n = \braket{\psi_{tar}|G_{tar}^{n-1}G|\psi_{in}}
    .
\end{equation}
This new operator is non-Hermitian but much like the original $G$ all of its eigenvalues are real and less than or equal to one.
Similarly to the way we used $G$'s eigenbasis we'll expand the initial probability vector $G\ket{\psi_{in}}$ using $G_{tar}$'s eigenbasis. We'll denote $G_{tar}$'s eigenvalues as $\mu$ and its right and left eigenstates as $\ket{\mu_{R}}$ and $\bra{\mu_{L}}$ where $G_{tar}\ket{\mu_{R}}=\mu \ket{\mu_{R}}$ and $\bra{\mu_{L}} G_{tar} = \bra{\mu_{L}} \mu$, as usual the left and right eigenvectors have the same eigenvalues \cite{leftrighteigenvalues}.
With these we can express the first detection probability in a manner similar to what we did in Eq. (\ref{eq:inanddetexpression1}), with a minor complication made by the fact that rather than span $\ket{\psi_{in}}$ using this eigenbasis we need to span $G\ket{\psi_{in}}$ instead, where we assume tat $G_{tar}$ has no degeneracies.
The aforementioned expansion gives us an equation for the first detection probability:
\begin{equation}
    \label{eq:fn3}
    F_n=\sum_{\mu} \mu^{n-1} \frac{\braket{\mu_{L}|G|\psi_{in}}}{\braket{\mu_{L}|\mu_{R}}} \braket{\psi_{tar}|\mu_{R}}
    .
\end{equation}
Eq. (\ref{eq:fn3}) is a formal solution of the problem.
It shows that $F_n$ decays exponentially as a function of $n$ provided that the sampling rate is non-exceptional.

Using Eq. (\ref{eq:fn3}) and the insight from Eq. (\ref{eq:pdetequation}) that the total detection probability for non-exceptional sampling rates is one, we can derive the following equation for the survival probability $S_N = 1 - \sum_{j=1}^{N} F_j$ which is the probability that in a finite system the particle was not detected, namely that it survived.
We find:
\begin{equation}
\label{eq:snequationfinal}
    S_N = \sum_{\mu} \frac{e^{N \ln{\mu}}}{1-\mu} \frac{\braket{\mu_{L}|G|\psi_{in}}}{\braket{\mu_{L}|\mu_{R}}} \braket{\psi_{tar}|\mu_{R}}
    .
\end{equation}
Based on this equation we can see that we should expect the survival probability to exponentially decay to zero for non-exceptional sampling rates and that the decay rate should slow down considerably when one or more of $G_{tar}$'s eigenvalues approaches one since the decay rates are given by $\ln{\mu}$.
As we will now show using an interlacing technique \cite{Hartich_2019}, this happens close to every exceptional sampling rate.

To see this, we first note that $G_{tar}$'s eigenvalues satisfy the equation:
\begin{equation}
    \text{Eigenvalues}(G_{tar}) = \{0\} \cup \text{Eigenvalues}(G')
\end{equation}
where $G'$ is the principle submatrix of $G$ where the row and column corresponding to $\psi_{tar}$ have been removed.
Note that $G'$ is Hermitian and also a principle submatrix of $G_{tar}$.
This is significant thanks to the Cauchy Interlace Theorem \cite{2005math2408F} which states that:

For any pair of Hermitian matrices $A$ and $B$ of order $N$ and $N-1$ respectively where $B$ is a principle submatrix of $A$ the eigenvalues of the two matrices interlace.
This means that if we label $A$'s eigenvalues as $a_n$ such that $a_{n-1} \leq a_n$ and label $B$'s eigenvalues as $b_n$ such that $b_{n-1} \leq b_n$ then:
\begin{equation}
    a_1 \leq b_1 \leq a_2 \leq b_2 \leq ... \leq b_{N-2} \leq a_{N-1} \leq b_{N-1} \leq a_N
    .
\end{equation}

\begin{figure}[h]
  \centering
  \includegraphics[width=0.5\textwidth]{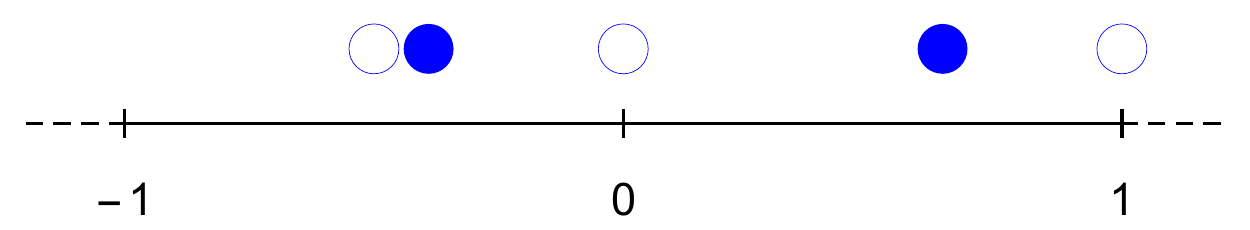}
  \caption{A simple illustration of the Eigenvalue interlacing theorem. The eigenvalues of $G$ and $G'$ for the Hamiltonian (\ref{eq:3sitelatticeHamiltonian}) plotted on a segment of the real number line. $G$s eigenvalues are depicted as blue circles whereas $G'$s eigenvalues are depicted as blue discs. We find that $G'$s eigenvalues always interlace between $G$s eigenvalues just as the theorem predicts.}
\label{fig:EigenvaluesOnTheLine}
\end{figure}

To show how this affects our problem using an example we examine a transition from $\ket{0}$ to $\ket{2}$ on a simple 3 site line graph which is described by the following tight binding Hamiltonian:
\begin{equation}
\label{eq:3sitelatticeHamiltonian}
    H = - \gamma \left( 
    \ket{0}\bra{1} + \ket{1}\bra{0} +
    \ket{1}\bra{2} + \ket{2}\bra{1}
    \right)
    .
\end{equation}
After using Eq. (\ref{eq:Gdefinition}) to find $G$ and $G'$, we evaluate them at the non-exceptional sampling rate $\gamma \tau = \pi/\sqrt{8}$ and then calculate their eigenvalues.
We find that the eigenvalues of the two matrices interlace just like the theorem predicts, as is shown in Fig \ref{fig:EigenvaluesOnTheLine}.

\begin{figure}[h]
  \centering
  \includegraphics[width=0.5\textwidth]{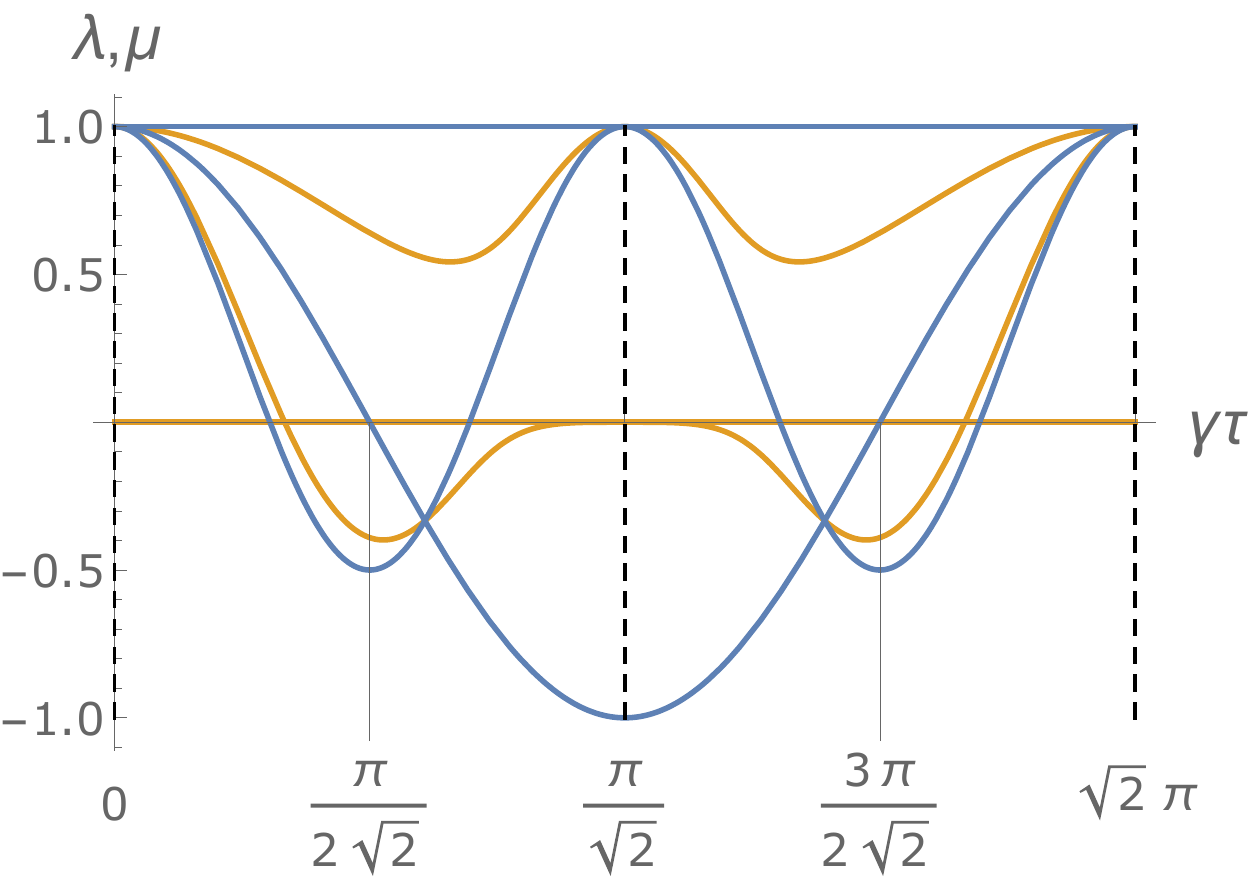}
  \caption{The eigenvalues of $G$ (denoted $\lambda$) and $G_{tar}$ (denoted $\mu$) plotted in blue and orange respectively.
  For all values of $\gamma \tau$ there is always a $\mu$ between every pair of $\lambda$ aside from zero which does not interlace.
  The eigenenergies of the system are $0 \text{, } \pm \sqrt{2} \gamma$.
  The exceptional sampling rates are $\gamma \tau = 0 \text{, } \pi/\sqrt{2} \text{, and } \sqrt{2}\pi$ plus integer multiples of $\sqrt{2} \pi$.
  In the figure these are marked by dashed lines, and all of them satisfy $\Delta E \tau = 2 \pi k$.
  Near each of the exceptional sampling rates we can see that at least one of the $\mu$s is squeezed between a pair of $\lambda$'s until it equals 1.}
  \label{fig:3sitelatticeEigenvalues}
\end{figure}

With the exception of zero (which is always an eigenvalue of $G_{tar}$), this interlacing occurs for all values of $\gamma\tau$, meaning that there is always a $\mu$ between every pair of $\lambda$s.
As $\tau$ approaches an exceptional sampling rate, one of $G$'s eigenvalues will approach unity.
Since $\lambda = 1$ is always an eigenvalue of $G$, the $\mu$ which is interlaced between it and the aforementioned eigenvalue must also coalesce on unity as it is squeezed between the two $G$ eigenvalues.
This causes the survival probability decay to slow down considerably since, as mentioned in Eq. (\ref{eq:snequationfinal}), $\mu \to 1$ implies slow relaxation. 
To show this we plot the eigenvalues of $G$ and $G_{tar}$ together as a function of $\gamma\tau$ in Fig. \ref{fig:3sitelatticeEigenvalues}.
We can see that near every exceptional sampling point, as one of $G$s eigenvalues approaches one, the $G_{tar}$ eigenvalue trapped between it and one is forced to also converge to one, in order to preserve the interlacing property.

\begin{figure}[h]
    \centering
    \includegraphics[width=0.5\textwidth]{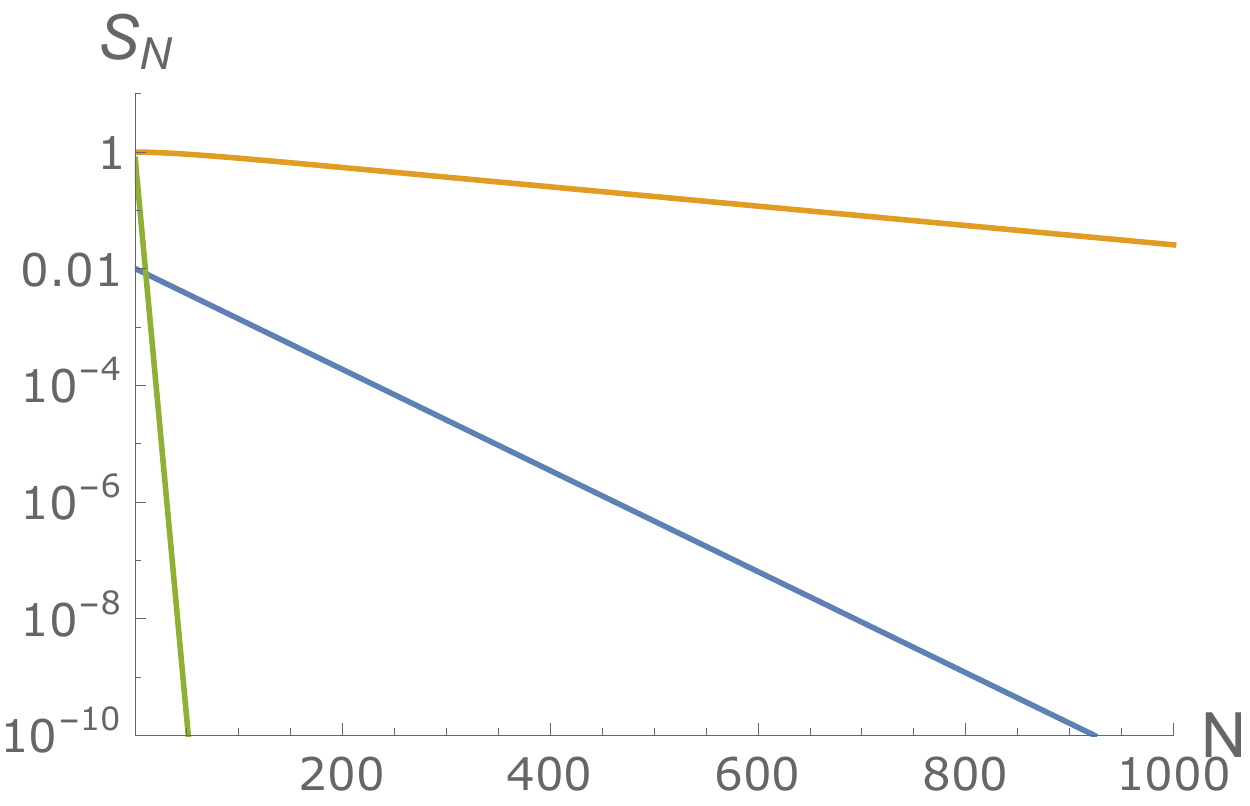}
    \caption{The survival probability in the transition from $\ket{0}$ to $\ket{c}$ on the graph whose Hamiltonian is described in Eq. (\ref{eq:3sitelatticeHamiltonian}). The orange and blue lines correspond to the almost exceptional sampling rates $\gamma \tau = \epsilon$ and $\pi / \sqrt{2} + \epsilon$ where $\epsilon=10^{-1}$. The green line corresponds to the non-exceptional sampling rate $\pi / \sqrt{8}$. As explained in the text the interlacing theorem predicts the slow decay of the survival probability close to the exceptional sampling rates through the analysis of the eigen values of G. When ever two eigen values of G merge we get a slow relaxation of the survival probability.}
    \label{fig:survivaldecayfig}
\end{figure}

To see how this affects the survival probability, we plot it as a function of $N$ for the two almost exceptional sampling rates $\gamma \tau = \epsilon$ and $\pi / \sqrt{2} -\epsilon$ where $\epsilon = 10^{-1}$.
We also plot along side them the survival probability for the non-exceptional sampling rate $\gamma \tau = \pi / \sqrt{8}$ for comparison.
As can be clearly seen in Fig. \ref{fig:survivaldecayfig} the decay rate is considerably slower for the two almost exceptional sampling rates compared to the non-exceptional one.

To summarize, the slow decay of the survival probability near certain sampling rates can be understood using the eigenvalue interlacing theorem.
We do this by searching for sampling rates that cause an eigenvalue of $G$ to approaches unity (besides $\ket{\phi}$s eigenvalue which is always one).
In general this can be done using any parameter of the Hamiltonian, not just the sampling rate.
This in turn implies that an eigenvalue $\mu$ will also approach unity and hence using Eq. (\ref{eq:fn3}, \ref{eq:snequationfinal}) the relaxation rate is reduced drastically unless prefactors vanish as well.
This analysis using $G_{tar}$ is in some senses redundant as we've already analyzed $G$.
However, in examining both matrices and their properties we are able to gain various insights into the behavior of the system which would be more difficult to notice in analyzing just one or the other.

\section{Benzene-like ring}
\label{sec:examples}
\begin{figure}[h]
  \centering
  \includegraphics[width=0.5\textwidth]{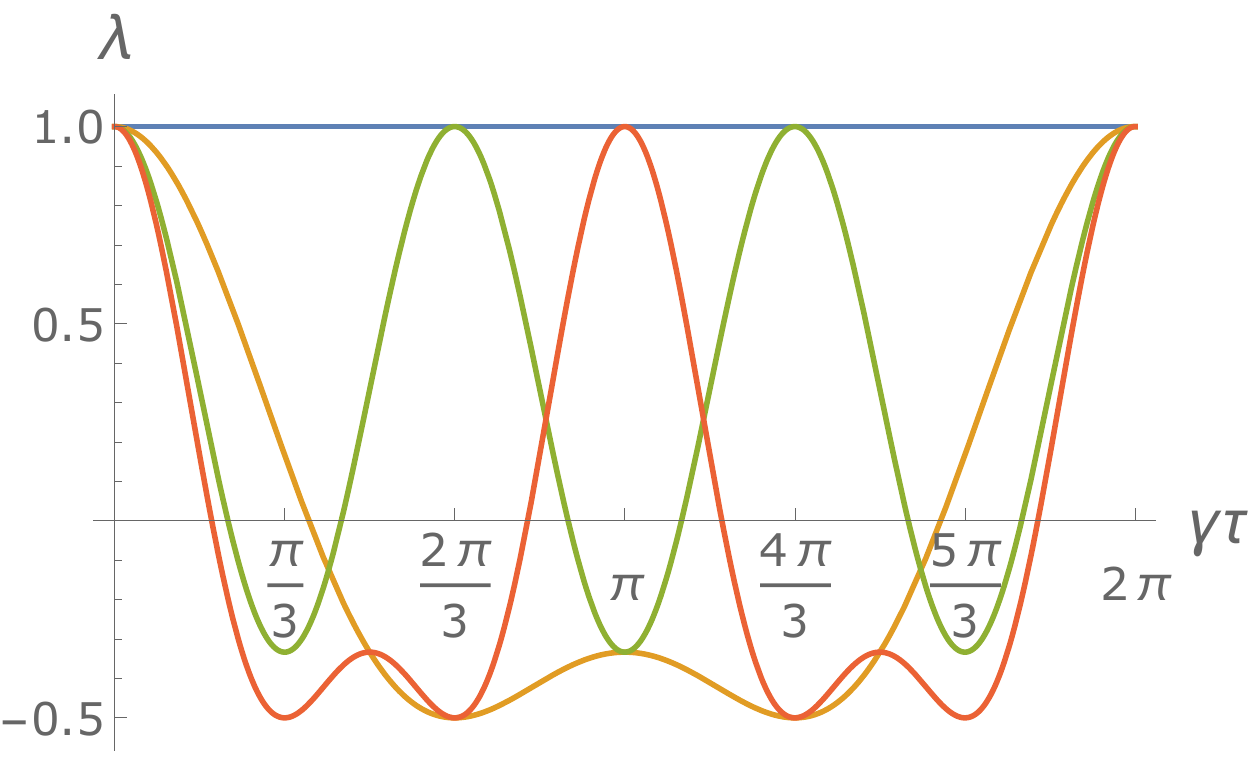}
  \caption{The eigenvalues of $G$ versus $\gamma \tau$ for the Benzene-like ring. As with every other system one of the Eigenvalues is a constant one while the others are sinusoidal functions of $\gamma \tau$. Interesting effects which are shown in the following figures are observed when $\lambda=1$ becomes degenerate. The exceptional sampling rates of this Hamiltonian are: $\gamma \tau = 0 \text{, } 2 \pi / 3  \text{, } \pi \text{, } 4 \pi / 3  \text{, } 2 \pi$ plus every integer multiple of $2 \pi$. These sampling rates all satisfy $\Delta E \tau = 2 \pi k$, where $\Delta E$ are the non-negative differences between eigenenergies of the Hamiltonian given in Eq. (\ref{eq:6ringH}). The eigenvalues plotted in orange and green have 2-fold degeneracy each.}
  \label{fig:HexagonEigenValues}
\end{figure}

As an example of how our formalism can be used we study the measurement induced quantum walk, and solve the return problem as well as a transition problem for a 6 site ring graph (see Fig \ref{fig:Hexagon}).
In this solution we omit some of the simpler steps of the process so as to instead focus on the results and their implications, in appendix \ref{appendix:twolevelsystem} we solve the first detection problem for a two level system giving the full technical details.
The graph is described by the following tight binding Hamiltonian with cyclical boundary conditions ($\ket{6}=\ket{0}$):
\begin{equation}
\label{eq:6ringH}
    H = -\gamma \sum_{x=0}^{5} \left(\ket{x}\bra{x+1} + \ket{x+1}\bra{x}\right)
    .
\end{equation}
We diagonalize the Hamiltonian to obtain $G$ and then diagonalize it to obtain the eigenstates and eigenvalues, the latter of which we plot as a function of $\gamma \tau$ in Fig \ref{fig:HexagonEigenValues}.

Having found $G$'s eigenstates and eigenvalues, the latter of which we plotted as a function of $\gamma \tau$ in Fig. \ref{fig:HexagonEigenValues}, we can now easily solve any measurement induced quantum first detection problem on the hexagonal graph.
Before we do that however, we first look at the time evolution of the probability vector free of absorbing boundary conditions $\ket{\rho(\tau n)}$, which we can easily obtain using:
\begin{equation}
    \ket{\rho(\tau n)} = \sum_{\lambda} \sum_{k=1}^{g_{\lambda}} \lambda^n \braket{\lambda_k|\psi_{in}}\ket{\lambda_k}
    .
\end{equation}
Doing this, we find that the projection of the probability vector unto every graph site equals $1/6$ plus an exponentially decaying sum of sinusoidal functions of $\gamma \tau$, meaning that in the long time limit the system decays to the state $\ket{\phi}$ as expected.
We also find that at exceptional sampling rates, some of said sinusoidal functions equal zero.
We provide an explanation of this phenomena and its physical implications near the end of this section.

Returning to the topic of the first detection problem, in this section we'll solve the return problem from $\ket{\psi_{in}} = \ket{0}$ to $\ket{\psi_{tar}} = \ket{0}$ and the transition problem from $\ket{\psi_{in}} = \ket{0}$ to $\ket{\psi_{tar}} = \ket{3}$.
Note that the choice to examine the transition problem from $\ket{0}$ to $\ket{3}$ as opposed to other less symmetric transitions is arbitrary and similar effects are observed for $\ket{0}$ to $\ket{1}$ and $\ket{0}$ to $\ket{2}$ as well.
We briefly address the other transitions at the end of this section.

\begin{figure}[h]
  \centering
  \includegraphics[width=0.5\textwidth]{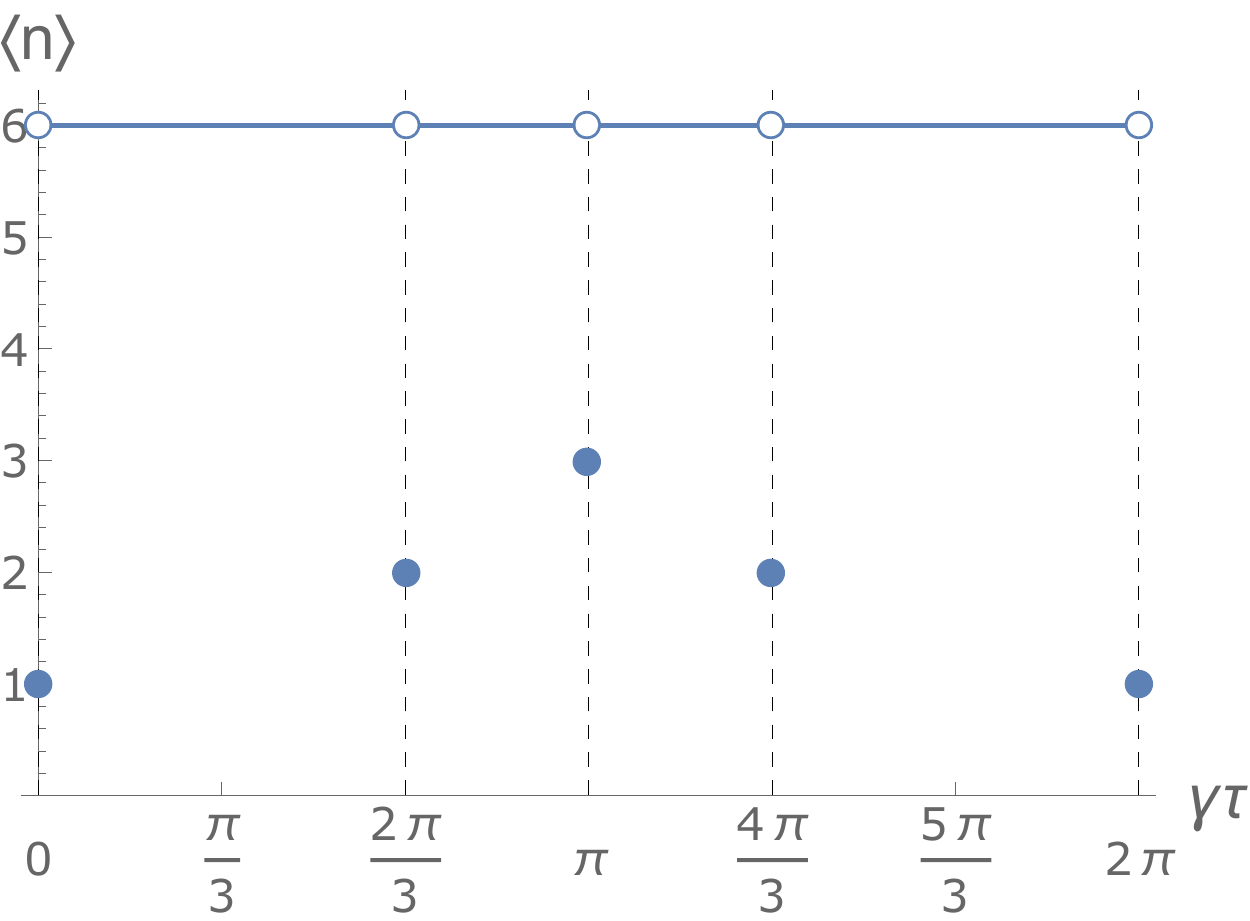}
  \caption{$\braket{n}$ for the return problem for a model ring with six sites whose Hamiltonian is given by Eq. (\ref{eq:6ringH}). As predicted by Eq. (\ref{eq:avgnfinalform2}) $\braket{n} = 6$ for non-exceptional sampling rates. We find that at every exceptional sampling rate $\braket{n}$ discontinuously jumps to a different integer smaller than six. In the text we show that this happens because at those values of $\gamma \tau$ the connectivity of the graph is broken, which separates it into several fragmented graphs.}
  \label{fig:HexagonReturnAverage}
\end{figure}

\begin{figure}[h]
  \centering
  \includegraphics[width=0.5\textwidth]{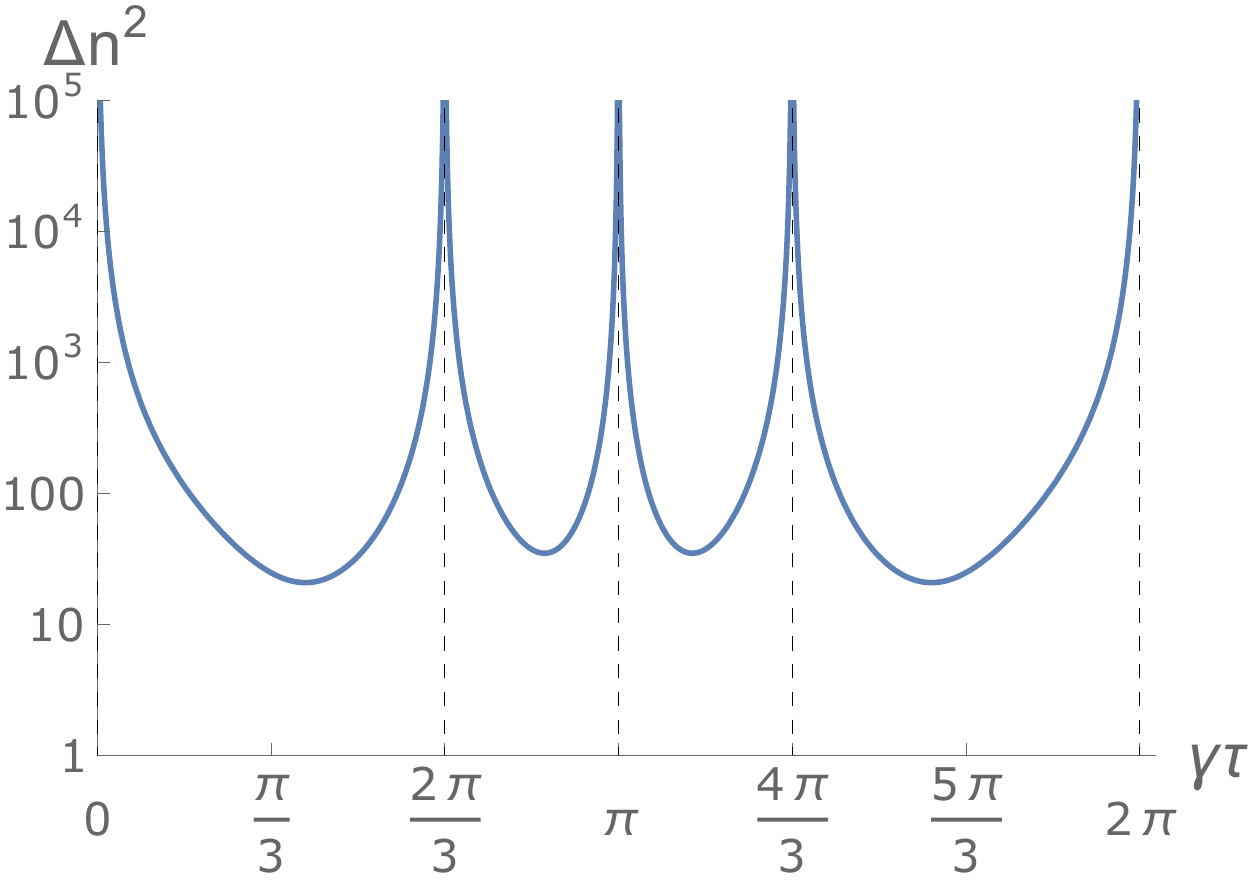}
  \caption{$\Delta n^2 = \braket{n^2} - \braket{n}^2$ for the return problem on a model ring with six sites whose Hamiltonian is given by Eq. (\ref{eq:6ringH}).
  As predicted in Eq. (\ref{eq:varianceformula}), we find that the variance diverges at every exceptional sampling rate. Namely, whenever a pair of lambda's in Fig \ref{fig:HexagonEigenValues}, coalesce on unity.}
  \label{fig:HexagonReturnVariance}
\end{figure}

\begin{figure}[h]
  \centering
  \includegraphics[width=0.5\textwidth]{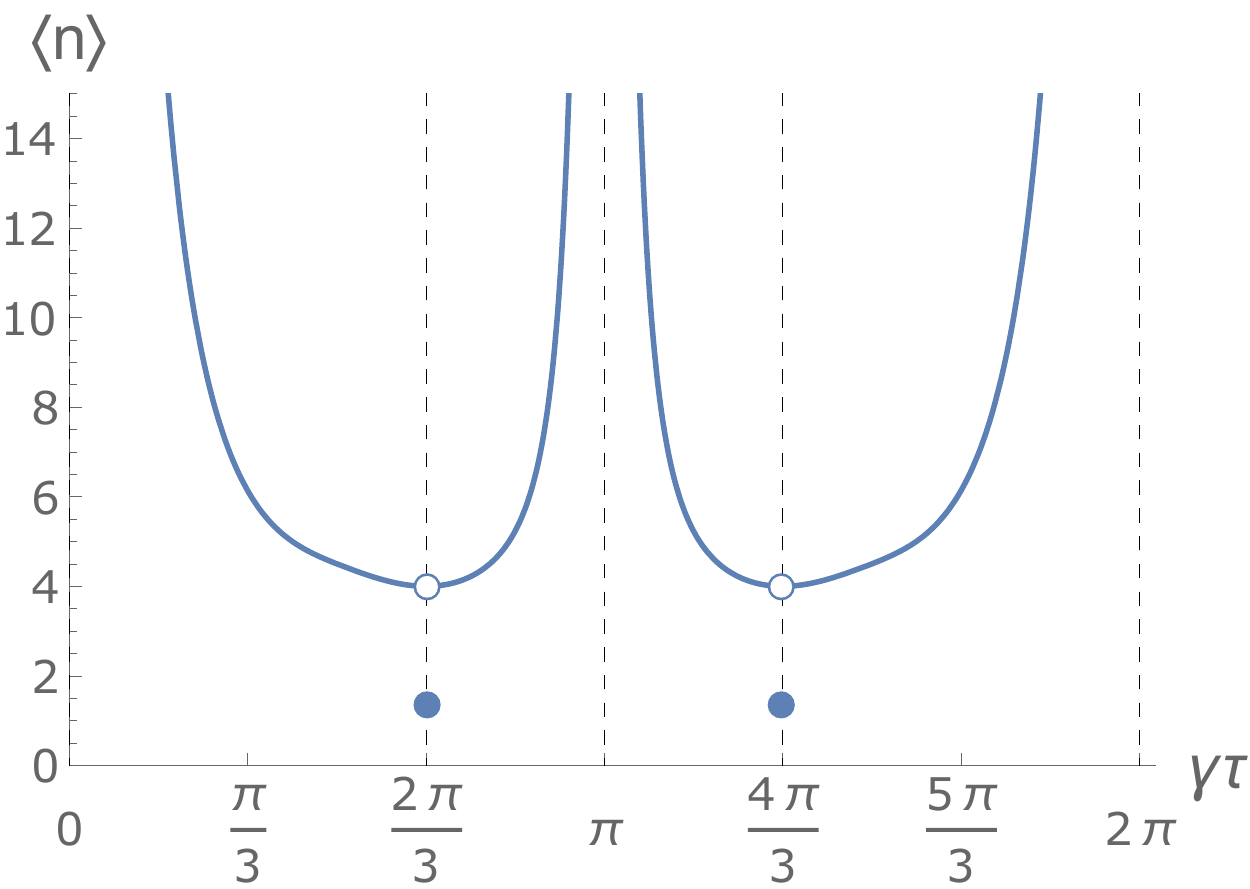}
  \caption{$\braket{n}$ for the transition problem from $\ket{\psi_{in}} = \ket{0}$ to $\ket{\psi_{tar}} = \ket{3}$ on a model ring with six sites whose Hamiltonian is given by Eq. (\ref{eq:6ringH}). In the text we explain why it diverges at $\gamma \tau = \pi n$ but only discontinuously jumps at the other sampling rates.}
  \label{fig:HexagonTransitionAverage}
\end{figure}

\begin{figure}[h]
  \centering
  \includegraphics[width=0.5\textwidth]{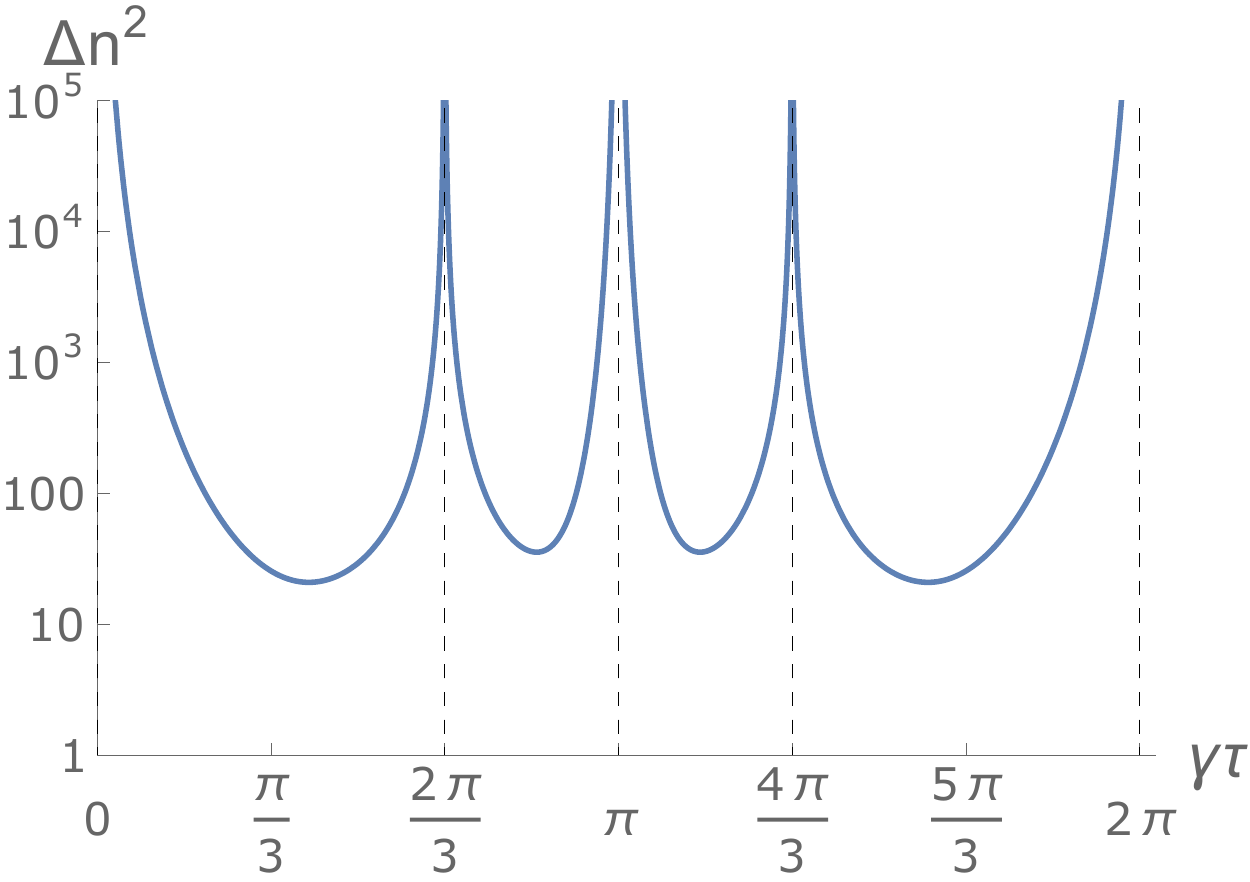}
  \caption{$\Delta n^2 = \braket{n^2} - \braket{n}^2$ for the transition problem from $\ket{\psi_{in}} = \ket{0}$ to $\ket{\psi_{tar}} = \ket{3}$ on a model ring with six sites whose Hamiltonian is given by Eq. (\ref{eq:6ringH}). We find that the variance diverges at every exceptional sampling rate, including $2 \pi n / 3$ in spite of the fact that the average does not diverge there and only discontinuously jumps.}
  \label{fig:HexagonTransitionVariance}
\end{figure}

In both the return and all of the possible transitions we find that the total detection probability $P_{det}$ is one for non-exceptional sampling rates. In the return problem, we find that $\braket{n}$ is discontinues at the exceptional sampling rates and in the transition problem we find that it often diverges. In both we find that the variance diverges at these sampling rates. In Figs. \ref{fig:HexagonReturnAverage} - \ref{fig:HexagonTransitionVariance} we plot these values as a function of $\gamma \tau$ and denote the exceptional values of $\gamma \tau$ with dashed vertical lines.


\begin{figure}[h]
     \centering
     \begin{subfigure}{0.115\textwidth}
         \centering
         \includegraphics[width=1\textwidth]{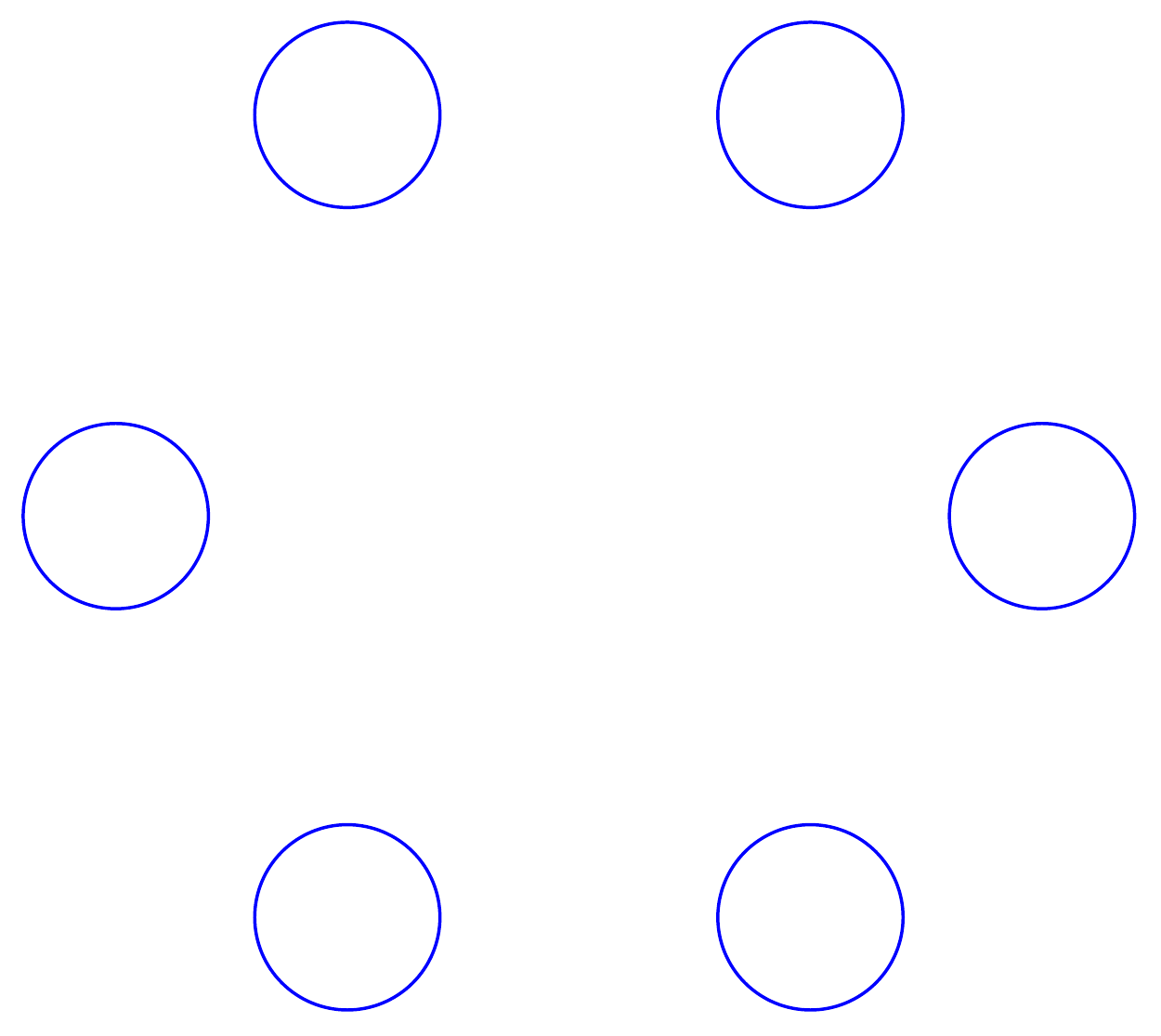}
         \caption{$\gamma \tau = 2 \pi n$}
         \label{fig:4hexs1}
     \end{subfigure}
     \begin{subfigure}{0.115\textwidth}
         \centering
         \includegraphics[width=1\textwidth]{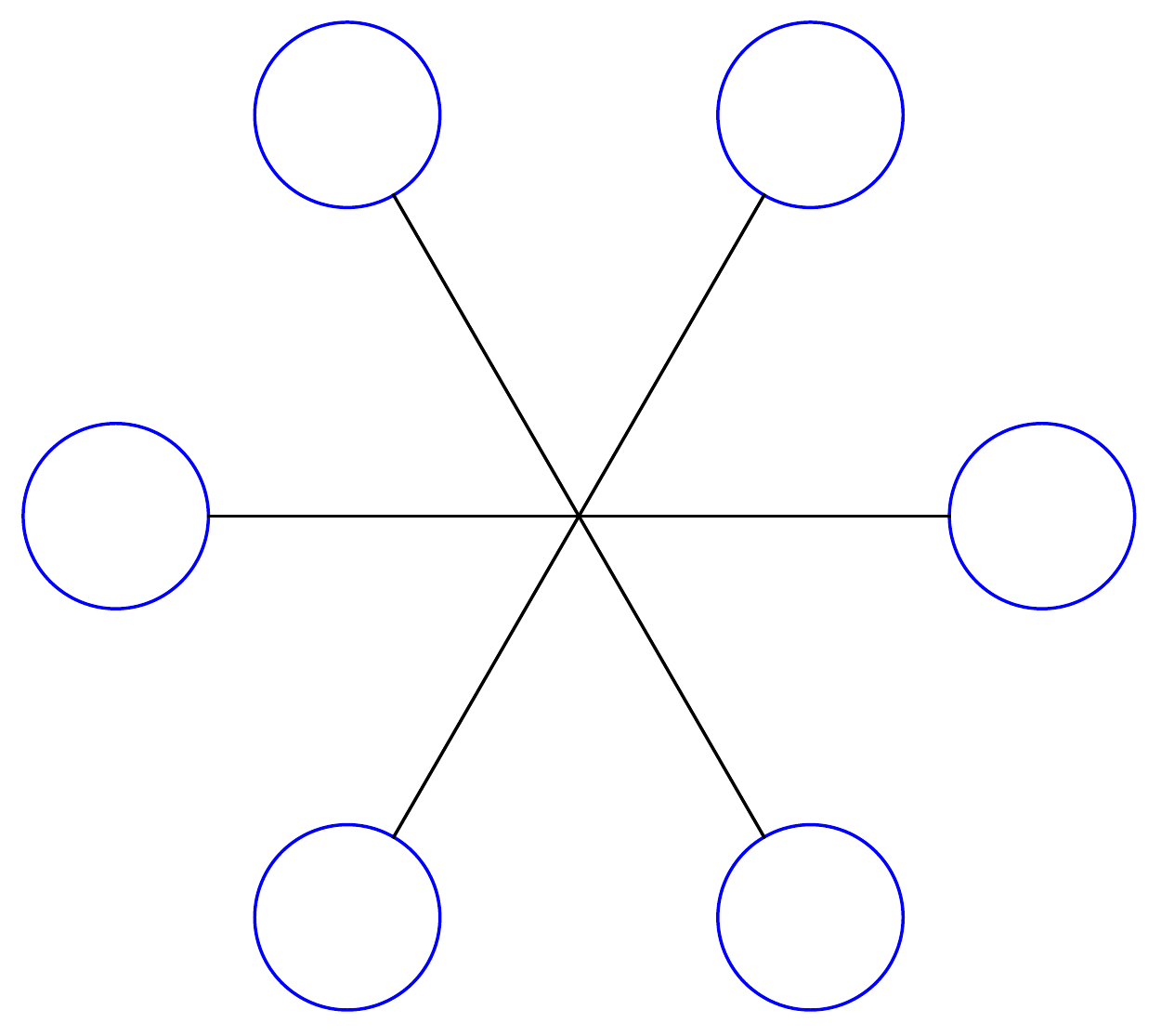}
         \caption{$\gamma \tau = \frac{2 \pi n}{3}$}
         \label{fig:4hexs2}
     \end{subfigure}
     \begin{subfigure}{0.115\textwidth}
         \centering
         \includegraphics[width=1\textwidth]{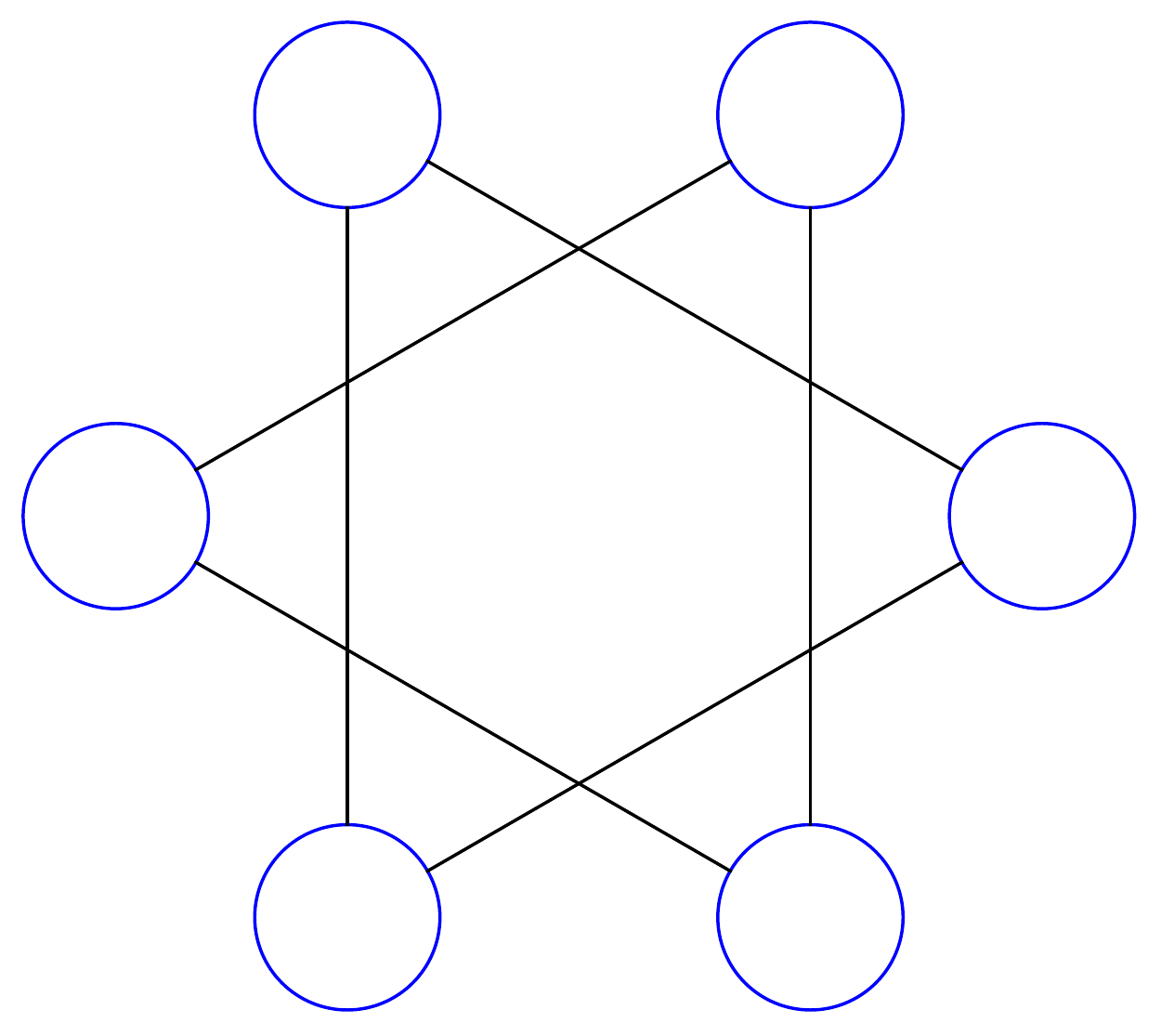}
         \caption{$\gamma \tau = \pi n$}
         \label{fig:4hexs3}
     \end{subfigure}
     \begin{subfigure}{0.115\textwidth}
         \centering
         \includegraphics[width=1\textwidth]{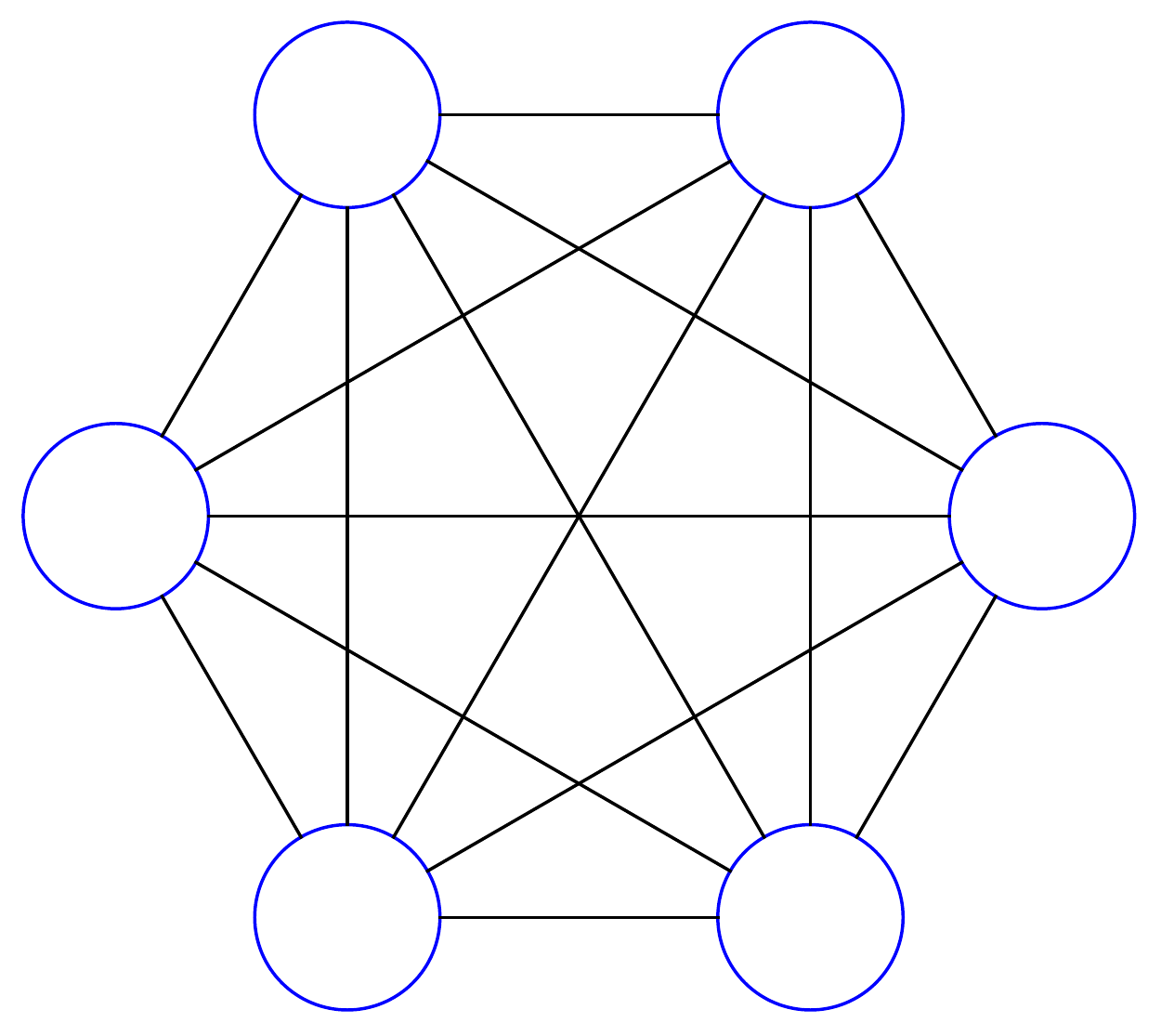}
         \caption{\scriptsize{Any other $\tau$}}
         \label{fig:4hexs4}
     \end{subfigure}
     \caption{The possible transitions for various sampling rates on a model ring with six sites whose Hamiltonian is given by Eq. (\ref{eq:6ringH}). In the text we explain how this result relates to the behavior of the system and in particular $P_{det}$ and $\braket{n}$. Note that in Fig $c$ $n$ is odd.}
     \label{fig:4hexsTABLE}
\end{figure}

In order to better understand these discontinuities and divergences we evaluate the stochastic matrix describing the process $G$ at the exceptional values of $\gamma \tau$ and observe that at those points the ergodicity of the system is broken.
In Fig. \ref{fig:4hexsTABLE}, we graph the system for various values of $\gamma \tau$ based on the $G$ we obtain for the sampling rate, this shows us what possible transitions exist in the system at these sampling rates.
We can see that the exceptional sampling rates are ones where the graph "breaks apart" into several disconnected sub-graphs, whereas for non-exceptional sampling rates every transition is possible with some non-zero probability.
This effect is somewhat similar to fragmentation of the Hilbert space \cite{thiel2019quantum, darkStatesPaper2, darkStatesPaper3} and the formation of Quantum Many-Body Scars \cite{QManyBodyScars, QManyBodyScarsLong}, and results in analogous behavior of the system.
Note that in Fig. \ref{fig:4hexsTABLE} the lines denote non-zero transition probabilities after a measurement, unlike in Fig. \ref{fig:Hexagon} where the lines denoted interactions between pairs of adjacent sites.

Looking at these graphs, we can also clearly see that the reason $\braket{n}$ in the transition problem diverged at $\gamma \tau = \pi$, but only discontinuously jumped to a different finite value at $\gamma \tau = 2 \pi n / 3$ is that whereas for $\gamma \tau = 2 \pi n / 3$ the point $\ket{3}$ remains reachable from $\ket{0}$ for $\gamma \tau =\pi n$ it becomes completely unreachable as demonstrated in Fig. \ref{fig:4hexsTABLE}.
Using Fig. \ref{fig:4hexsTABLE}, we can also see that in the return problem the average is still equal to the size of the graph even for exceptional sampling rates and the reason it discontinuously jumps from six to smaller integer is that the graph itself is effectively split apart at the exceptional sampling rates.
For example for $\gamma \tau = 2 \pi n$, the effective size of the graph is not six but rather one (as shown in Fig. \ref{fig:4hexs1}).
Hence for these sampling times, $\braket{n} = 1$ for the return problem.

Based on Fig. \ref{fig:4hexsTABLE}, we can also see that similar behaviour will occur for other transitions.
In the transition $\ket{0}$ to $\ket{1}$, $\braket{n}$ will diverge at every exceptional sampling rate, since $\ket{1}$ becomes unreachable at all of them.
Whereas in the transition from $\ket{0}$ to $\ket{2}$, $\braket{n}$ will diverge at $\gamma \tau = 2 \pi n / 3$, since $\ket{2}$ becomes unreachable at this value and it will discontinuously jump at $\gamma \tau = \pi$.

Looking back at the probability vector $\ket{\rho(\tau n)}$, we can also see that as expected at exceptional sampling rates rather than decay to $\ket{\phi}$ which is evenly distributed across the entire graph it decay to an even distribution across only the sub-graph $\ket{\psi_{in}}$ belongs to.

\section{Unbounded quantum walker in one dimension}
\label{sec:infiniteline}
In this section we consider a measurement induced quantum walk for a free particle on an infinite lattice.
In classical random walk theory this is the problem of first passage time for a particle diffusing without bias on a lattice in one dimension, which is of course a well studied problem \cite{SidRedner}.

We use the tight-binding Hamiltonian:
\begin{equation}
    \label{eq:inflineH}
    H = -\gamma\sum_{x=-\infty}^{\infty} \ket{x}\bra{x + 1} + \ket{x + 1}\bra{x}
    .
\end{equation}
A schematic description of this system is given in Fig \ref{fig:inflinegraphix}.

We start by first examining the behavior of the probability vector without absorbing boundary conditions, which is given by Eq. (\ref{eq:probvecwithoutabsorbingboundaryconditions}).
The solution of the Schr\"{o}dinger equation for the Hamiltonian (\ref{eq:inflineH}) is $\ket{\psi(t)} = \sum_{x=-\infty}^{\infty}B_x\ket{x}$ where the amplitudes satisfy $i \dot{B}_x=-\gamma (B_{x+1}+B_{x-1})$. Using the Bessel function identity $2J'_v(z)=J_{v-1}(z)-J_{v+1}(z)$ \cite{HandbookOfMath} and the initial condition $B_x(t=0)=\delta_{x,0}$ we find that without measurement, the wave function is:
\begin{equation}
    \ket{\psi(t)} = \sum_{x=-\infty}^{\infty}i^x J_x(2\gamma t)\ket{x}
    .
\end{equation}
From this it is clear that the time evolution operator $G$ will be:
\begin{equation}
    \label{eq:inflinetimeevomatrix}
    G = \sum_{x, x'=-\infty}^{\infty} |J_{x-x'}(2\gamma \tau)|^2 \ket{x}\bra{x'}
    .
\end{equation}
Using this we find that the Fourier transform of the probability vector at time $t=\tau n$ is given by:
\begin{equation}
\label{eq:probvecFourierTransform}
    \widetilde{P}_{k,n} = \sum_k e^{-ikx} \braket{x|\rho(\tau n)} = J_0(4 \gamma \tau \sin(\frac{k}{2}))^n
\end{equation}
We present the exact derivation of this expression in appendix \ref{app:inflinederivationappendix}.
We can obtain the moments of the random walk using derivatives of this Fourier transform.
The first moment $\braket{x}$ is zero from symmetry and the second moment which corresponds to the unitless variance in position of the random walk is:
\begin{equation}
    \Delta x^2 = 2 n \gamma^2 \tau^2
    .
\end{equation}

\begin{figure}[H]
     \centering
     \begin{subfigure}{0.42\textwidth}
         \centering
         \includegraphics[width=1\textwidth]{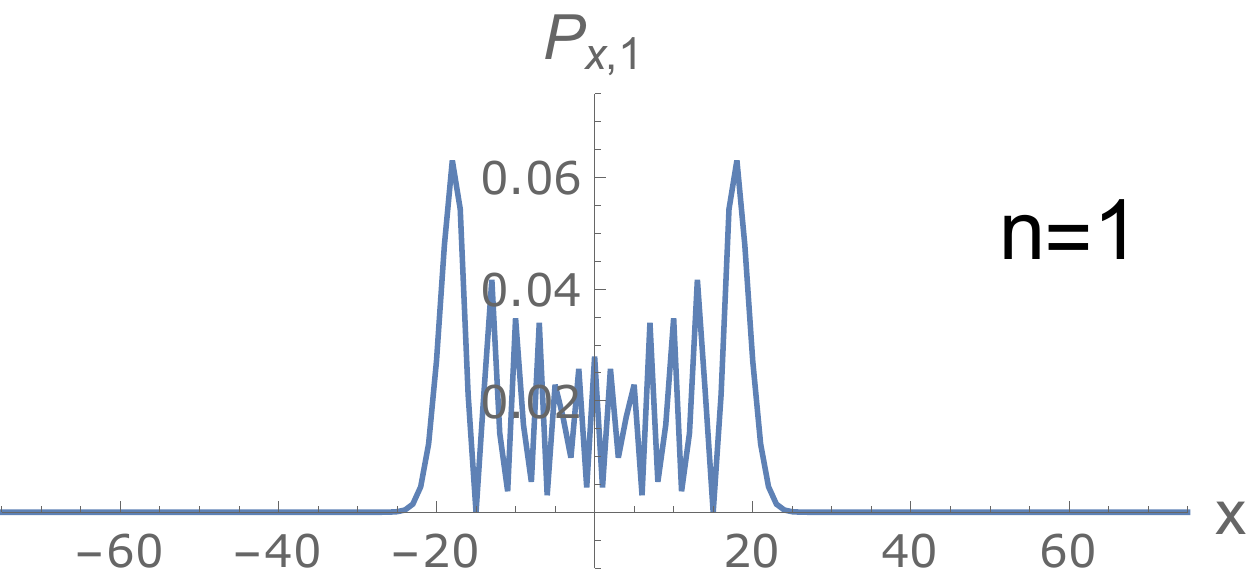}
     \end{subfigure}
     \begin{subfigure}{0.42\textwidth}
         \centering
         \includegraphics[width=1\textwidth]{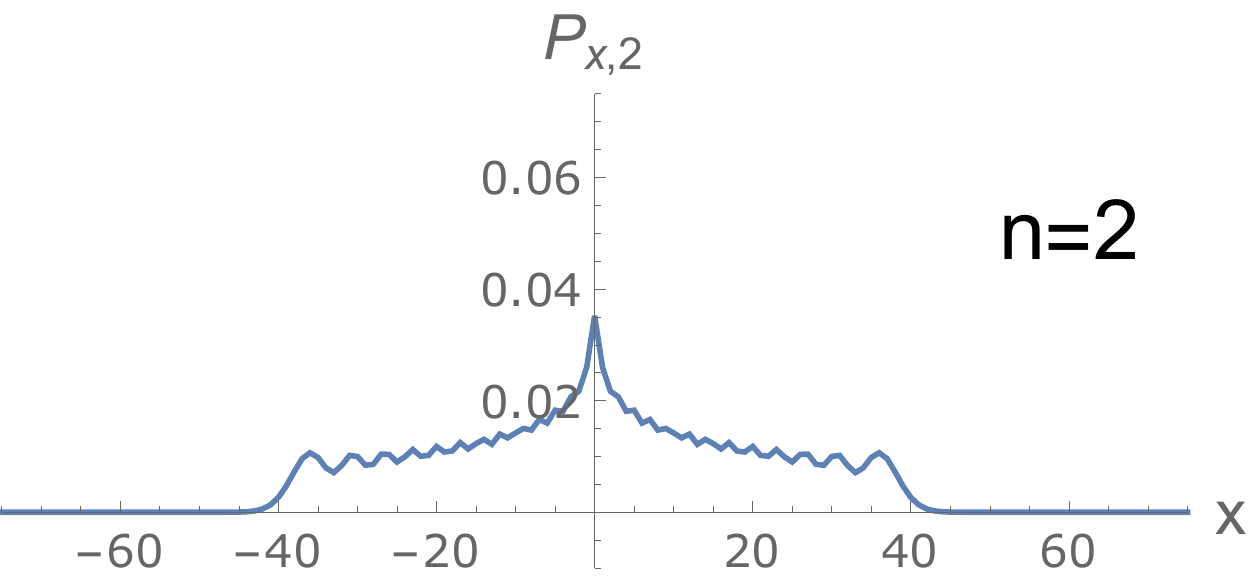}
     \end{subfigure}
     \\
     \begin{subfigure}{0.42\textwidth}
         \centering
         \includegraphics[width=1\textwidth]{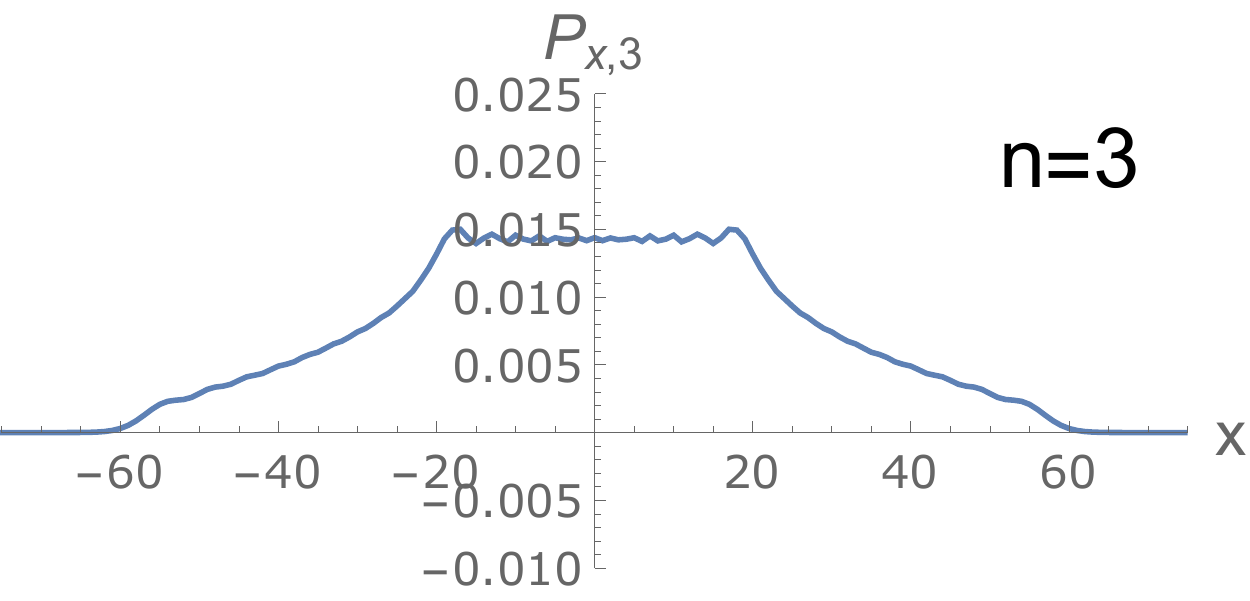}
     \end{subfigure}
     \begin{subfigure}{0.42\textwidth}
         \centering
         \includegraphics[width=1\textwidth]{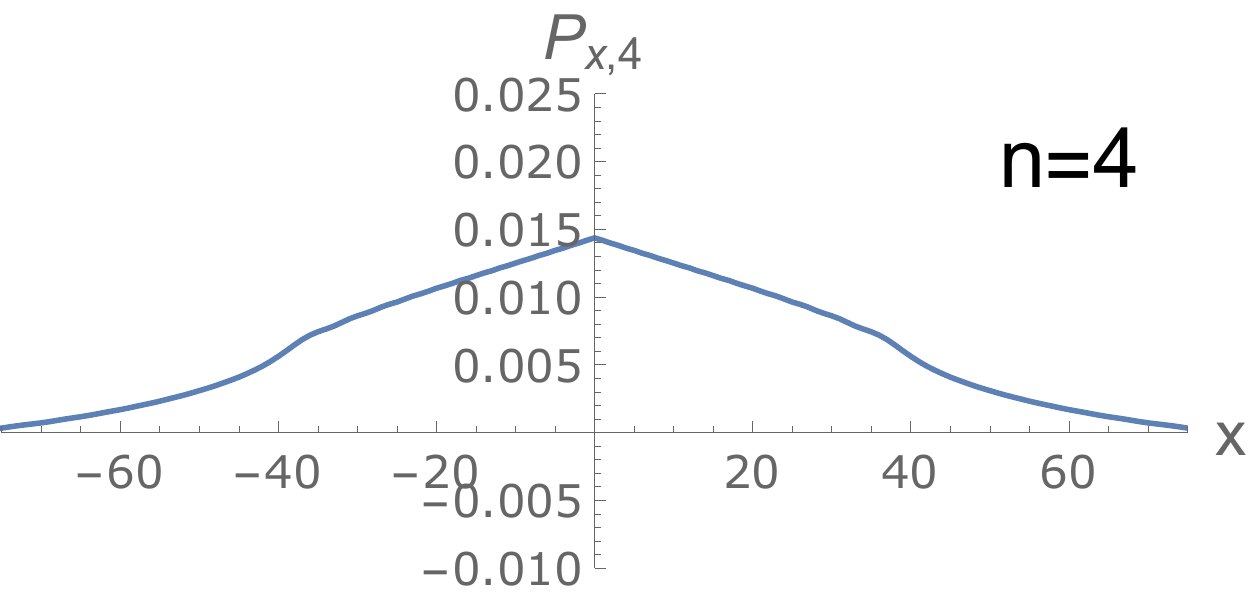}
     \end{subfigure}
     \caption{
The probability vector on the infinite line Hamiltonian given in Eq. (\ref{eq:inflineH}) after the first four measurements plotted as a function of the lattice site $x$ with $\gamma \tau = 10$.
The shape after the first measurement matches the results obtained in experimental implementations of this type of random walk \cite{PhysRevLett.100.170506}, indicating that the Hamiltonian we use (Eq. (\ref{eq:inflineH})) can accurately model such a system.
We tested many values of the sampling rate and found that, excluding $\gamma \tau \ll 1$, the general shape of the distribution after the first few measurements is the one presented here.
The scale of the shape depends on the sampling rate, and we find that the "width" equals $2 n \gamma \tau$ as can be seen in this figure.
More generally, we can say that the maximum group velocity of the wave packet when evolving without measurement is $2 \gamma$, and this is reflected here in the distance that we can detect the particle in.
From the fifth measurement onward the probability vector typically converges to a simple Gaussian shape near the center as expected, we study the behavior of the tails of the distribution separately later in this section.
}
\label{fig:inflineprobvecFIRSTFOUR}
\end{figure}

The probability vector after the first four measurements is shown for reference in Fig. \ref{fig:inflineprobvecFIRSTFOUR}.

\subsection{Edgeworth series}
We'll start by examining the region of the random walk close to the origin.
As one might expect, in this region the probability vector quickly converges to a Gaussian, which can be shown with a cumulant expansion up to the second order.
\begin{equation}
    J_0(4 \gamma \tau \sin(\frac{k}{2}))^n \sim e^{- n \gamma^2 \tau^2 k^2}
\end{equation}
Under this simple approximation the probability vector equals:
\begin{equation}
\label{eq:gaussianexpressionone}
    \braket{x|\rho(\tau n)} = P_{x,n} \approx \frac{\exp (-\frac{x^2}{4 n \gamma^2 \tau^2})}{\sqrt{4 \pi n \gamma^2 \tau^2}}
\end{equation}
We can reintroduce $\hbar$ to Eq. (\ref{eq:gaussianexpressionone}) and add a lattice size $a$ which we define to be the distance between different sites on the lattice to get:
\begin{equation}
    P_{x,n} \approx \frac{\exp (-\frac{\hbar^2 x^2}{4 n a^2 \gamma^2 \tau^2})}{\sqrt{4 \pi n a^2 \gamma^2 \tau^2 / \hbar^2}}
\end{equation}

In order to find the corrections to the Gaussian near the origin we'll use an Edgeworth series \cite{10.2307/2339343}, which is given by:
\begin{equation}
\label{EqEdgeworthSeriesFullExpression}
    P_{x,n} = \frac{\exp{(-\frac{x^2}{2 c_2})}}{\sqrt{2 \pi c_2}} \left( 1 + \sum_{l=2}^{\infty} 
    \frac{(-1)^l c_{2l}}{(2l)! (2c_2)^l} H_{2l}(\frac{x}{\sqrt{2 c_2}})
    \right)
\end{equation}
where $H_m(x)$ are the Hermite polynomials and $c_l$ is the $l$th cumulant of the random walk, these can be obtained using:
\begin{equation}
    c_l = i^l \lim_{k \to 0} \frac{d^l}{dk^l} \ln{(\widetilde{P}_{k,n})} = i^l n \lim_{k \to 0} \frac{d^l}{dk^l} \ln{(\widetilde{P}_{k,1})}
    .
\end{equation}
Since the cumulants are linearly proportional to n, the corrections to the Gaussian in Eq. (\ref{EqEdgeworthSeriesFullExpression}) disappear over time as expected.
A short list of some of the cumulants is given in Table \ref{tableCumulants}.

\begin{figure}[h]
  \centering
  \includegraphics[width=0.45\textwidth]{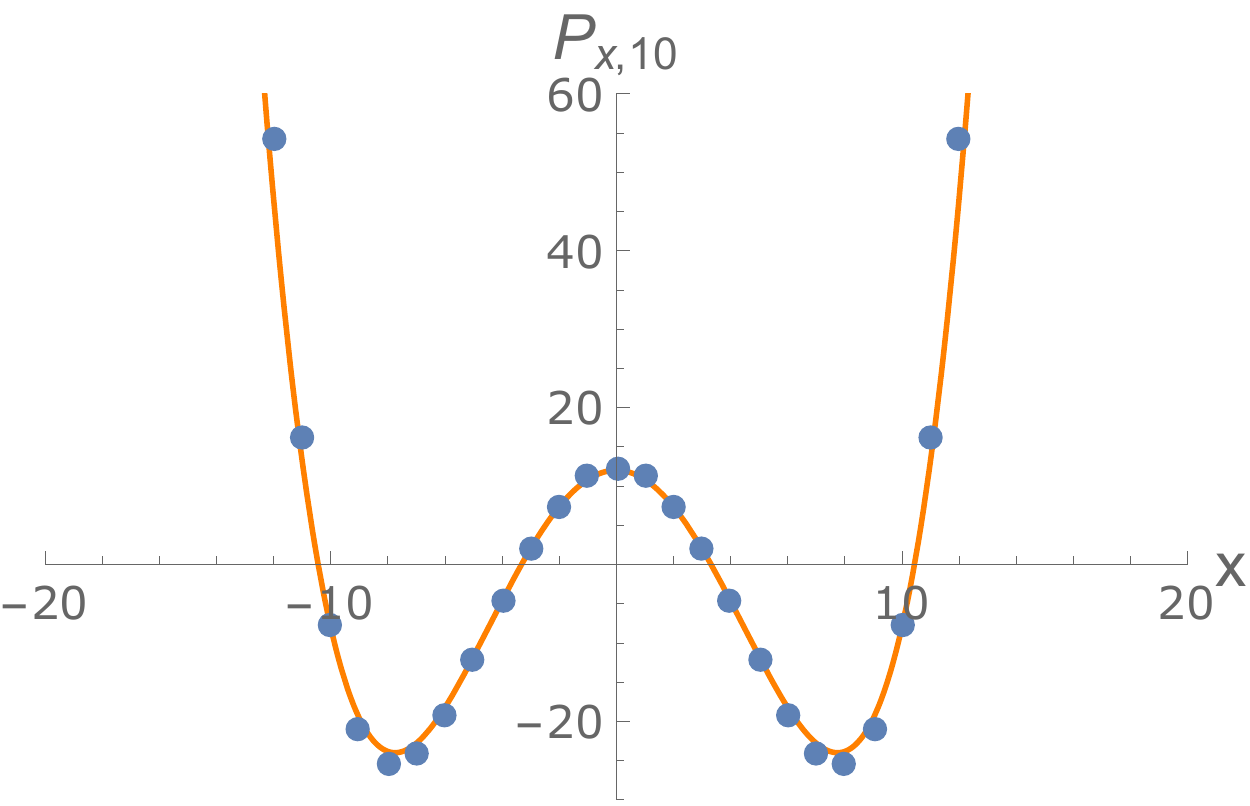}
  \caption{A plot of the expression given in Eq. (\ref{EqApproximationForFourthHermitPoly}) for $\gamma \tau = 1$ and $n=10$.
  The probabilities of detection which were obtained via numerical integration are plotted as blue dots and the 4th Hermite polynomial is plotted as an orange line.
  Numerical simulations indicate that typically the two overlap almost completely for $10 \leq n$.
  }
  \label{figInfLineHemitPoly4}
\end{figure}

\begin{table}[H]
\centering
  \begin{tabularx}{0.48\textwidth} { 
  | >{\centering\arraybackslash}c
  | >{\raggedright\arraybackslash}X | }
\hline
$n$ &  $c_n$ \\ 
\hline
$0$ & $ 0 $ \\ 
\hline
$2$ & $2 n \gamma ^2 \tau ^2$ \\ 
\hline
$4$ & $2 n \gamma ^2 \tau ^2 \left(1-3 \gamma ^2 \tau ^2\right)$ \\ 
\hline
$6$ & $2 n \gamma ^2 \tau ^2 \left(1 - 15 \gamma ^2 \tau ^2 + 40 \gamma ^4 \tau ^4\right)$ \\ 
\hline
$8$ &  $2 n \gamma ^2 \tau ^2 \left(1 - 63 \gamma ^2 \tau ^2 + 560 \gamma ^4 \tau ^4 - 1155 \gamma ^6 \tau ^6 \right)$ \\ 
\hline
$10$ & \footnotesize{$2 n \gamma ^2 \tau ^2 \left(1 - 255 \gamma ^2 \tau^2 + 5880 \gamma ^4 \tau ^4 - 34650 \gamma ^6 \tau ^6 + 57456 \gamma ^8 \tau ^8\right)$} \\ 
\hline
  \end{tabularx}
\caption{The first ten Cumulants of the random walk on the infinite line, which are the coefficient of the series expansion of the natural log of Eq. (\ref{eq:probvecFourierTransform}) at $k=0$. It should be noted that the special value of $\gamma \tau$ which causes the Kurtosis $c_4 / \sigma^4$ to go to zero has no particular significance with regard to the Gaussianity of the distribution, and is just coincidental.}
\label{tableCumulants}
\end{table}
The Edgeworth series gives us a more accurate approximation of the probability vector near the origin than just the Gaussian, which is it's first term.
One way to see this is that we can actually retrieve the Hermite polynomials from the random walk itself.
In order to obtain the 4th Hermite polynomial, which is the second term of the series, we can divide $P_{x,n}$ by the Gaussian term to get:
\begin{equation}
\label{EqApproximationForFourthHermitPoly}
    H_{4}(\frac{x}{\sqrt{2 c_2}}) \approx \frac{(4)! (2c_2)^2} {c_{4}} (\frac{P_{x,n}}{N_{x,n}} - 1)
\end{equation}
where $N_{x,n}$ is a Gaussian ($N$ormal distribution) with $\mu = 0$ and $\sigma^2 = c_2$.
In Fig. \ref{figInfLineHemitPoly4} we demonstrate this using a numerical simulation of the random walk.

\subsection{Saddle point}
\label{secSadlePoint}
\begin{figure}[h]
  \centering
  \includegraphics[width=0.45\textwidth]{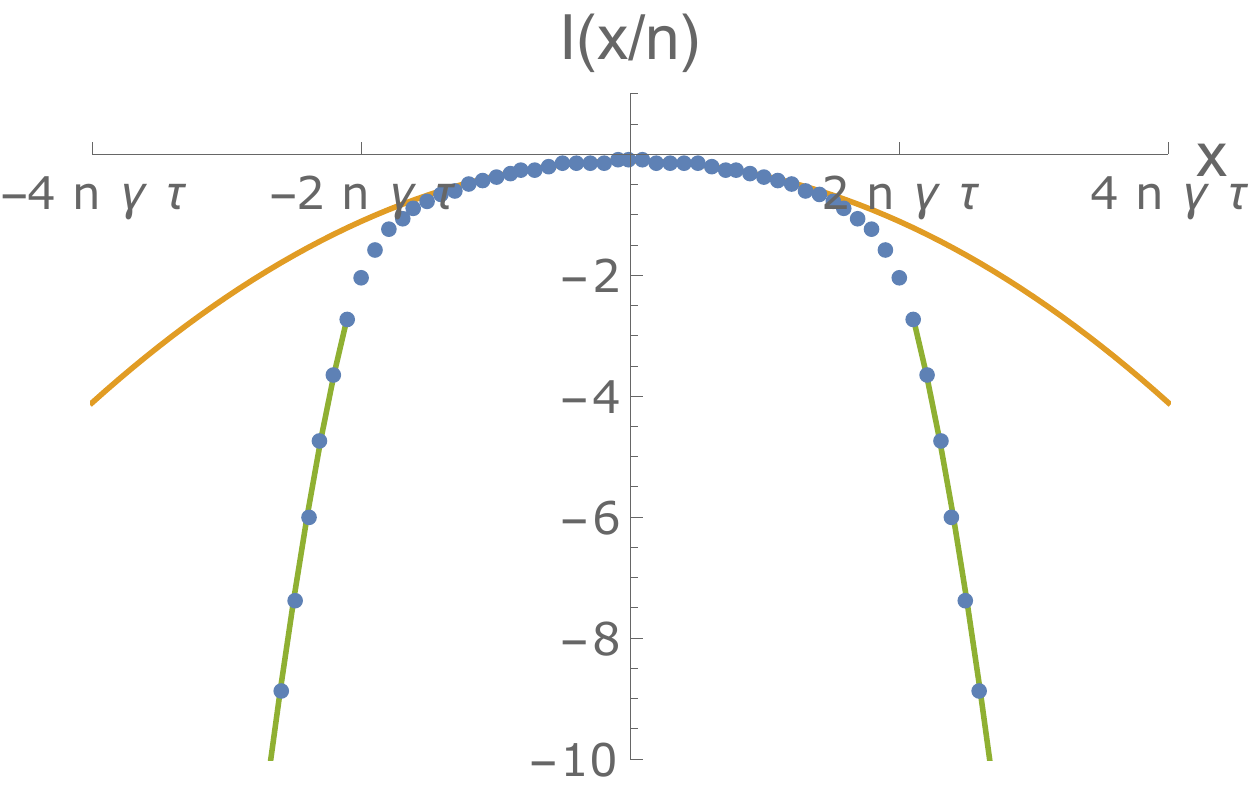}
  \caption{
  The rate function $I(x/n) = n^{-1} \ln (P_{x,n})$ on the infinite 1D line plotted in blue as a function of the lattice site $x$.
  The small $l$ approximation (Gaussian) is plotted in orange and the large $l$ approximation (Eq. (\ref{eq:bigLapproxinthething})) is plotted in green.
  We tested many values of the sampling rate and found that, excluding $\gamma \tau \ll 1$, the distribution fits the Gaussian approximation up to around $|x| = 2 n \gamma \tau$ and then almost immediately fits the large $l$ approximation.
  }
  \label{figRateFunctionRender}
\end{figure}

While the Edgeworth series is useful for studying the probability vector near the origin, it's accuracy decreases drastically as we move towards the tails of the distribution, to the point that it can sometimes give negative values for the probabilities.
For these regions we use a saddle point method \cite{10.1214/aoms/1177728652} to approximate the probabilities for large values of $n$, we give the full details of the derivation in appendix \ref{appendixSaddlepoint}.
While our focus with this method is the tails of this distribution, it should be noted that using it to approximate the distribution near the origin gives a Gaussian distribution as expected (Appendix \ref{appendixsubsectionsmoll}).
For the far tails of the distribution, which we define as the region where $2 n \gamma \tau \ll x$, we find in appendix \ref{appendixsubsectionbigel} that the probabilities are given by:
\begin{equation}
\label{eq:bigLapproxinthething}
    P_{x,n} \approx \frac{1}{
    2 \pi n \gamma \tau \sqrt{\left( \frac{x}{2 \gamma \tau n} \right)^2 - 1}
    }\frac{
    I_0 \left( 4 \gamma \tau \sqrt{\left( \frac{x}{2 \gamma \tau n} \right)^2 - 1} \right)^n
    }{
    \exp \left( 2 x \text{ arccosh} \left( \frac{x}{2 \gamma \tau n} \right) \right)
    }
\end{equation}
Where $I_0(x)$ is the zeroth modified Bessel function of the first kind.
In the region after $x = 2 n \gamma \tau$ we find that the probabilities decay rapidly, much faster than a Gaussian, as shown in Fig. \ref{figRateFunctionRender}.

\subsection{Zeno limit}
Another topic of interest in this random walk is the Zeno limit, where we take $\tau \to 0$.
We can gain some simple insight into the behavior of the probability vector in this limit by taking the limit of $G$, which gives us:
\begin{equation}
    G \approx \sum_{x=-\infty}^{\infty} (1 - 2 \gamma^2 \tau^2) \ket{x}\bra{x} + \gamma^2 \tau^2 (\ket{x+1}\bra{x} + \ket{x}\bra{x+1})
\end{equation}
Since the prefactors of the probability of transition is quadratic in $\tau$, even if we consider the actual time $t = \tau n$ as opposed to the number of steps the particle is still forced to remain at the origin and has a very low probability of escaping.
Another way of seeing this, is that the diffusion coefficient of this random walk is proportional to $\tau^2$. 
We can better quantify this effect by examining the Kurtosis of the probability vector.
The Kurtosis is a measure of the Gaussianity of the system, it is zero for Gaussian distributions and expected to be small for distributions which are very similar to Gaussians.
In our previous discussion of the Edgeworth series we mentioned that we expect the probability vector to converge to a Gaussian fairly quickly, and using the Kurtosis we can quantify how quickly this happens.
The Kurtosis for this random walk is:
\begin{equation}
\label{EqKurtosis1}
    \kappa = \frac{c_4}{(c_2)^2} = \frac{(\gamma \tau)^{-2} -3}{2 n} \approx \frac{(\gamma \tau)^{-2}}{2 n}
\end{equation}
(The $-3$ is negligible in the Zeno limit).
To obtain a Kurtosis smaller than $\epsilon$ where $\epsilon \ll 1$, we can rearrange Eq. (\ref{EqKurtosis1}) to get:
\begin{equation}
\label{EqKurtosis2}
    \frac{1}{2} \epsilon^{-1} \leq n (\gamma \tau)^{2}
\end{equation}
Meaning that $n$ must be at least of the order of $(\gamma \tau)^{-2}$ in order for the probability vector to converge to a Gaussian shape.

\subsection{First detection time}
Turning our attention to the first detection time for the return problem on this lattice, we can easily obtain the generating function by plugging the inverse Fourier transform of Eq. (\ref{eq:probvecFourierTransform}) into Eq. (\ref{eq:GeneratingFunctionProbVecVersion2RETURN}) and perform the summation with respect to $n$ to get:
\begin{equation}
    \label{eq:InfLineGenFunc}
    \widetilde{F(z)} = 1 - \frac{2\pi}{\int_{0}^{2\pi}\frac{dk}{1-zJ_0(4 \gamma \tau \sin(\frac{k}{2}))}}
    .
\end{equation}

\begin{figure}[h]
  \centering
  \includegraphics[width=0.5\textwidth]{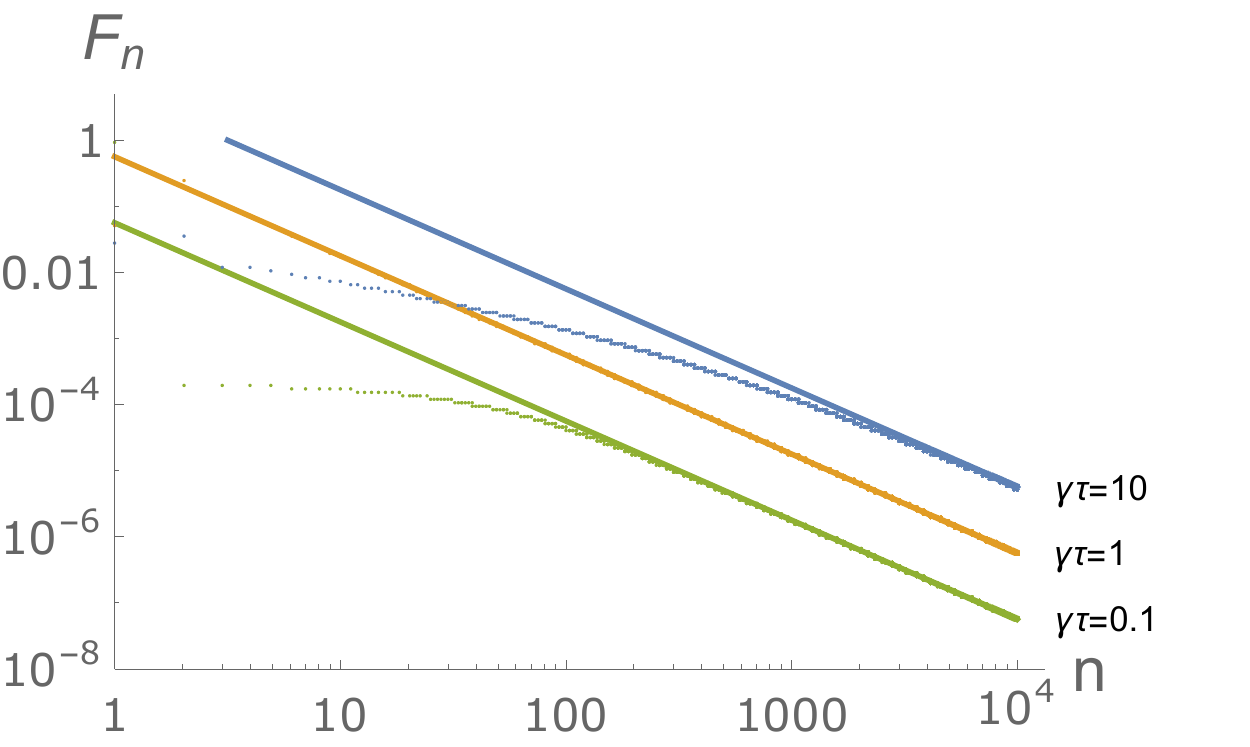}
  \caption{$F_n$ of the return problem on the infinite 1D line for various values of $\gamma \tau$ plotted as dots alongside their respective asymptotic limits, which were plotted as lines. The choice of sampling rate has a strong effect on the first few first detection probabilities, but after those they all converge to $\pi^{-0.5} \gamma \tau n^{-3/2}$. This convergence is generally faster for smaller values of $\gamma \tau$ but it slows down close to the Zeno limit.}
  \label{fig:InfiniteLineFnPlot}
\end{figure}

Since the integral in Eq. (\ref{eq:InfLineGenFunc}) diverges for $z \to 1$, using Eq. (\ref{<n^0>}) we know that for all $\tau$, $P_{det}=1$.
The small $n$ first detection probabilities can be obtained using Eq. (\ref{eq:Fn2}) or Eq. (\ref{eq:F_nEq}) by expanding Eq. (\ref{eq:InfLineGenFunc}) around $z=0$ and numerically solving the resulting integrals.
The exact form of the first few of these are presented in Table \ref{table:firstfewinflinefn}.
We can also continue with such a process with a program like Mathematica to obtain $F_n$, which we plot versus $n$ in Fig. \ref{fig:InfiniteLineFnPlot}.

\begin{table}[ht]
\centering
  \begin{tabularx}{0.45\textwidth} { 
  | >{\centering\arraybackslash}c
  | >{\raggedright\arraybackslash}X | }
\hline
$n$ &  $F_n$ \\ 
\hline
$1$ & $ A_1 = |J_{0}(2\gamma \tau)|^2$ \\ 
\hline
$2$ & $A_2 - A_1^2$ \\ 
\hline
$3$ & $ A_3 - A_1 A_2 + A_1^3 $ \\ 
\hline
$4$ &  $ A_4 + 3 A_1^2 A_2 -2 A_1 A_3 - A_2^2 - A_1^4 $ \\ 
\hline
$5$ & $ A_5 + 3 A_1^2 A_3 + 3 A_1 A_2^2 - 2 A_1 A_4 - 4 A_1^3 A_2 - 2 A_2 A_3 + A_1^5 $ \\ 
\hline
  \end{tabularx}
\caption{The first 5 first detection probabilities for the return problem on the infinite line lattice.
In this table we denoted $(2\pi)^{-1} \int_{0}^{2 \pi} (J_0 (4 \gamma \tau \sin{(k/2)}) )^n dk$ as $A_n$ for the sake of readability.}
\label{table:firstfewinflinefn}
\end{table}

In order to obtain the asymptotic behaviour of the probabilities for large n we can solve the integral in Eq. (\ref{eq:InfLineGenFunc}) in the limit $1-z \ll 1$. In this limit, most of the contribution to the integral is in the region of $k=0$ so we'll expand the Bessel function around this point up to the second order $J_0(4 \gamma \tau \sin (k)) \approx 1 - 4 \gamma^2 \tau^2 k^2$.
This approximation is accurate in the region which contributes the most to the integral and quickly decays to zero is the region which by comparison contributes very little.
\\
With this approximation we obtain:
\begin{equation}
    \widetilde{F(z)} \approx 1- 2 \gamma \tau \sqrt{1-z}
    .
\end{equation}
From the expansion of the function about $z=1$ using the Tauberian theorem \cite{SidRedner} we can see that the asymptotic big n behaviour is:
\begin{equation}
    F_{1 \ll n} \approx \pi^{-1/2} \gamma \tau n^{-3/2}
    .
\end{equation}
The asymptotic $n^{-3/2}$ behavior is of course similar to well known result from one dimensional first passage times (classical) on a line \cite{SidRedner}.
We show the convergence of the first detection probabilities to this limit in Fig. \ref{fig:InfiniteLineFnPlot}.
The prefactor we find here is the new element of this research.

In addition to the example we've considered here, an examination of the effects of Anderson localization on this kind of random walk would be interesting, as the combination of the localization caused by the measurements with the localization of the Hamiltonian eigenstates caused by disorder in the lattice might result in unique effects.

\section{Comparison with target site measurement}
\label{section:comparison}
One last subject to consider is alternate definitions for a quantum first detection problem.
In this paper we considered a quantum first detection problem where every $\tau$ time we measure the position of the particle until it reaches the target site.
One other possible definition of a quantum first detection problem is one where every $\tau$ time we just measure if the particle reached the target site or not using the projection operator $\ket{\psi_{tar}}\bra{\psi_{tar}}$ as opposed to the position operator $\hat{X}$.
This issue was considered in \cite{PhysRevE.95.032141, PhysRevResearch.2.033113, PhysRevResearch.1.033086}.
In the latter case the outcome of each measurement is either yes or no, hence we can claim that the measurement is local.
The basic renewal equations Eq. (\ref{eq:Fn2}) here and Eq. (17) in \cite{PhysRevE.95.032141} are fundamentally different, as here we describe a probability $F_n$ and the latter an amplitude of first detection.
In the  classical world the two protocols give the same first passage time statistics, but in quantum mechanics this is not the case.
The current problem is more classical, but some traces of quantum mechanics are still present in the features of G, such as the exceptional sampling rates, the Zeno limit, etc.

For a more quantitative examination, we compare some of the general results obtained for both measurement protocols in order to highlight the differences between the two.
Of particular interest are $P_{det}$ and $\braket{n}$ for a finite Hilbert space, since we have general results for both in both measurement protocols that apply for all time independent Hamiltonians at non-exceptional sampling rates. Note that in general the exceptional sampling rates in the position measurement protocol and the target site measurement protocol are different, though some can be the same (such as the Zeno limit, which exists in every system and for both protocols).

For the return problem, we found in this paper that the total detection probability is always one.
Similarly in the case when measurements are performed only at the target site, which in this case is also the initial location of the wave function.
The total detection probability is one in a finite Hilbert space with a discrete energy spectrum \cite{RecurrenceArticle}.
We found that $\braket{n}$ equals the size of the graph in our measurement protocol, whereas when only measuring at the target site it was found that $\braket{n}$ equals the effective dimension of the Hilbert space, which is the number of distinct energy levels of the system, provided that the corresponding eigenstates have a non-zero overlap with the detected state.
This effective dimension is always less than or equal to the size of the graph, so the target site measurement protocol results in faster detection on average.
Note that this is true typically in ordered system where we have some symmetry and hence degeneracy in the energy levels.
Whereas in a disordered systems the number of energy levels is typically the same as the dimension of the Hilbert space, causing the two averages to be the same.
Similar results are found also for non-Hermitian descriptions of the system \cite{PhysRevA.102.022210}.

For the transition problem, we found in this paper that the total detection probability is also always one, whereas for the target site measurement protocol $P_{det}$ can be bounded numbers smaller than unity depending on the symmetry of the system as was shown in \cite{thiel2019quantum, darkStatesPaper2, darkStatesPaper3}.
In general, for systems with symmetry the repeated local measurements yield states which are called dark states \cite{thiel2019quantum, darkStatesPaper2, darkStatesPaper3} and here we did not find such generic states (with the exception of the exceptional sampling rates, which are related to the stroboscopic protocol under study).

Although comparing $\braket{n}$ in the transition problem is not as easy as we don't have simple way of determining which measurement scheme is slower on average for it in general, we feel that the probability of not detecting the particle at all in the target site measurement scheme makes the location measurement scheme preferable for reliable detection of the particle in the transition problem.
However, for specific target states, the target site measurement approach can be extremely fast, as the motion is essentially ballistic \cite{krovi1}.

\section{Summary}
\label{sec:summarysec}
In this paper we've developed a theoretical framework for the study of quantum walks with repeated measurements of the position operator, which we've named the measurement induced quantum walk.
In particular we've focused on studying the first detection problem within this framework.
We found that rather than study the behavior of the wave function directly which is made complicated by the mix of unitary time evolution and non-unitary measurement induced collapse, we can obtain a simpler view of the problem by analysing the spectral properties of the stochastic matrix describing our Markov chains transition probabilities $G$.
Using the renewal Eq. (\ref{eq:Fn2}), we are able to study any aspect of the first detection statistics.
We've shown that in finite systems close to some sampling rates which we call exceptional the decay of the survival probability slows down, and that at the exceptional sampling rates themselves the behavior of our system changes drastically.

One remarkable feature of the return problem in a finite graph we've found is that the mean $\braket{n}$ is quantized and equal to the size of the graph.
This breaks down at exceptional sampling rates when the effective size of the system is smaller than the actual size, due to ergodicity breaking (see Fig. \ref{fig:4hexsTABLE}).
Importantly, it remains an integer, and so does $\braket{n}$.
For the transition problem, we have a very different behavior as $\braket{n}$ is certainly not an integer.
Instead it typically diverges close to exceptional sampling rates, but not always, as can be seen in Fig. \ref{fig:HexagonTransitionAverage}.
Although the discontinuities in $\braket{n}$ (such as those seen in Fig. \ref{fig:HexagonReturnAverage}) are very difficult to measure directly as any slight change in the sampling rate or noise from the environment will ruin this exceptional sampling rate, we can still see the effect of these sampling rates in the divergence of the variance.
Hence to study the effects of exceptional sampling rates, one does not need to tune the system very precisely and may instead simply focus of the fluctuations. 

For a particle on a one dimensional infinite lattice, we found that $F_n$ decays like $n^{-3/2}$.
This is different from what was found in \cite{PhysRevE.95.032141} for the same Hamiltonian when measuring only at the origin where the decay rate was found to be $F_n \sim n^{-3}$.
Hence the value of the exponent of the first detection probabilities depends on the observable used to define the problem.
In our case, the exponent $3/2$ is the same as the one obtained for a regular classical random walk.
Of course this does not imply that the problem itself is classical, only that the exponent $3/2$ is.

\section{Acknowledgments}
The support of Israel Science Foundation's grant 1898/17 is acknowledged.
We thank David Kessler, Felix Thiel, Ruoyu Yin, and Quancheng Liu for discussions and comments.

\appendix
\section{Derivation of $F_n$}
\label{appendix:Fn}
In this appendix we present all of the steps leading of the derivation of Eq. (\ref{eq:Fn2}) in detail.
\subsection{The time evolution operator}
We'll start by showing that using $G$ as the time evolution operator of the probability vector does in fact cause it to behave as we've described in Sec. \ref{sec:probabilityvector}.
We'll prove $G\ket{\rho(\tau n^+)} = \ket{\rho(\tau (n+1)^-)}$ by induction.

For $n=0$ we get $G\ket{\rho(0)} = \sum_x |\braket{x|e^{-i \tau H}|\psi_{in}}|^2 \ket{x}$, which is simply the probability that the wave function collapsed to each of those sites.
For $0<n$, we have $\ket{\rho(\tau n)} = \sum_x P_x^{\tau n} \ket{x}$ where by the induction hypotheses $P_x^{\tau n}$ is the probability that the wave function is localized to $\ket{x}$ at $t = \tau n$.
The probability that $\ket{\psi}$ is localized to some arbitrary site $\ket{y}$ at $t = \tau (n+1)$ is $\sum_x P_x^{\tau n} P_{x \to y}$ where $P_{x \to y}$ is the probability that a wave function starting at $\ket{x}$ collapses to $\ket{y}$ after being operated on by $e^{-i \tau H}$, which is given by $|\braket{y|e^{-i\tau h}|x}|^2$.
Operating on $\ket{\rho(\tau n)}$ with $G$ we get:
\begin{equation}
\label{appendix:G defining property}
    G\ket{\rho(\tau n)} = \sum_x \left[ \sum_{x'} P_{x'}^{\tau n} P_{x' \to x} \right] \ket{x} = \sum_x P_{x}^{\tau (n+1)} \ket{x}
    .
\end{equation}
$\sum_x P_{x}^{\tau (n+1)} \ket{x}$ equals $\ket{\rho(\tau (n+1))}$ by the definition of $P_n^t$, so the proof is complete.
\subsection{Evolution of the probability vector}
\label{appendix:fnEq1Derivation}
In this subsection we'll derive Eq. (\ref{eq:Fn1}).
Note that in this subsection we use $P_n$ to refer to the conditional probability that the wave function is localized to $\ket{\psi_{tar}}$ at $\tau n$ after having not been detected at $\psi_{tar}$ in the past $n-1$ attempts, as opposed to $F_n$ which is the probability of first detecting the particle at $\psi_{tar}$ on the nth measurement.
The relation between the two is $F_n = P_n \Pi_{j=1}^{n-1}(1-P_j)$. We'll also be making frequent use of the operator $D = \ket{\psi_{tar}}\bra{\psi_{tar}}$.

For $n=1$, Eq. (\ref{eq:Fn1}) follows directly from the defining property of G which was proven in the previous subsection of this appendix, and the probability that the wave function is localized to $\psi_{tar}$ is $F_1=P_1=\braket{\psi_{tar}|G|\psi_{in}}$.
If the system was not measured to be in the state $\psi_{tar}$, then at $t=\tau + \epsilon$ $(\epsilon \in \mathbb{R}^+, \epsilon \to 0^+)$ the probability vector is $\ket{\rho(\tau + \epsilon)} = N(1-D)\ket{\rho(\tau)}$ where N is the normalization of the probability, which in this case is $N = (1-P_1)^{-1}$. Continuing to evolve the vector in time, we find that $\ket{\rho(2 \tau)} = \frac{G(1-D)G\ket{\psi_{in}}}{1-P_1}$. If the system was not found at $\psi_{tar}$ again, then the probability vector is $\ket{\rho(2 \tau + \epsilon)} = \frac{(1-D)G(1-D)G\ket{\psi_{in}}}{(1-P_1)(1-P_2)}$.

This iteration procedure is repeated, with the operator $G(1-D)$ removing the component which was not detected at the target site and evolving the probability vector in time and the normalization $(1-P_j)^{-1}$ being added with each such failed detection attempt. After continuing in this manner for $\tau n$ time, we have:
\begin{equation}
    \ket{\rho(\tau n)}=\frac{(G(1-D))^{n-1}G\ket{\psi_{in}}}{\Pi_{j=1}^{n-1}(1-P_{j})}
    .
\end{equation}
Applying $\bra{\psi_{tar}}$ to both sides and multiplying by $\Pi_{j=1}^{n-1}(1-P_{j})$ we have:
\begin{equation}
    P_{n} \Pi_{j=1}^{n-1}(1-P_{j}) = \braket{\psi_{tar}|(G(1-D))^{n-1}G|\psi_{in}}
\end{equation}
The left hand side of this equation is the probability that the nth measurement attempt succeeded and all previous attempts failed, which is how we defined $F_n$. Hence, the derivation of Eq. (\ref{eq:Fn1}) is complete.

\subsection{Renewal equation derivation}
\label{appendix:equationEquivelance}
In this section we'll show by induction that
\begin{equation}
\label{appendix:F_n1}
    \left[ G ( \mathbb{1} - D)\right]^{n-1}G\ket{\psi_{in}} = 
    G^{n}\ket{\psi_{in}} - \sum_{j=1}^{n-1} F_{j} G^{n-j} \ket{\psi_{tar}}
    .
\end{equation}

For $n=1$, it's easy to see that both are just $G\ket{\psi_{in}}$ We'll now assume that the equation is correct for $n$ and prove that it follows that it's true for $n+1$.\\
Operating on the left hand side of Eq. (\ref{appendix:F_n1}) with $G(\mathbb{1} - D)$ we get $\left[ G ( \mathbb{1} - D)\right]^{n}G\ket{\psi_{in}}$.
Operating on the right hand side with $G(\mathbb{1} - D)$ we get:
\begin{equation}
\label{appendix: induction proof thing}
\begin{split}
    G^{n+1}
    &
    \ket{\psi_{in}} - \sum_{j=1}^{n-1} F_{j} G^{n+1-j} \ket{\psi_{tar}} -
    \\
    &
    \left[ \braket{\psi_{tar}|G^n|\psi_{in}} - \sum_{j=1}^{n-1} F_{j} \braket{\psi_{tar}|G^{n-j}|\psi_{tar}} \right] G\ket{\psi_{tar}}
    .
\end{split}
\end{equation}

Notice that the segment in square brackets in Eq. (\ref{appendix: induction proof thing}) is $F_n$. After applying this change and making the rightmost expression part of the sum we get:
\begin{equation}
\label{appendix:F_n2}
    \left[ G ( \mathbb{1} - D)\right]^{n}G\ket{\psi_{in}} = 
    G^{n+1}\ket{\psi_{in}} - \sum_{j=1}^{n} F_{j} G^{n+1-j} \ket{\psi_{tar}}
    .
\end{equation}

This equation is the same as (\ref{appendix:F_n1}) but for $n+1$. Hence, the proof is complete.
To get the equivalence of Eq. (\ref{eq:Fn1}) and (\ref{eq:Fn2}) from this, simply operate on Eq. (\ref{appendix:F_n2}) with $\bra{\psi_{tar}}$.

\section{Eigenvalues and eigenstates of $G$}
In order to complete the derivations of the general formulas for the moments of the generating function some general properties of $G$'s eigenvalues and eigenstates are necessary.
\label{appendix:eigenG}
\subsection{All Eigenvalues of $G$ are $\leq$ 1}
According to the Gershgorin circle theorem \cite{zbMATH03002670}, all of $G$'s eigenvalues lie within 'Gershgorin discs' $D(G_{ii}, R_{i})$ which are disks in the complex plane where $G_{ii}$ is the center of the disc and $R_{i} = \sum_{i \neq j}|G_{ij}|$ is the radius of the disc.
Since $G$ is Hermitian, we can replace the statement of the theorem with $\forall_{\lambda}\exists_{i}:|G_{ii} - \lambda| \leq R_{i}$ where $\lambda$ are the eigenvalues of $G$.
Since the total probability of a particle to jump to some other site from any initial site is one, we can replace $R_i$ with $1-G_{ii}$ and use that to rewrite the previous equation as follows: $\forall_{\lambda}\exists_{i}:|G_{ii} - \lambda| \leq 1 - G_{ii}$.
If $G_{ii} \leq \lambda$, it easily follows that $\forall |\lambda| \leq 1$.
Otherwise, if $\lambda < G_{ii}$, than it is also less than or equal to 1 since $\forall_iG_{ii}\leq1$.
\subsection{The vector $\ket{\phi} = \frac{1}{\sqrt{|X|}}\sum_{x}\ket{x}$ is an eigenstate of $G$}
\begin{equation}
    G\ket{\phi} = 
    \frac{1}{\sqrt{|X|}}\sum_{x}\left[ \sum_{x'} |\braket{x|e^{-iH\tau}|x'}|^2 \right] \ket{x} = \ket{\phi}
    .
\end{equation}
The term in the square brackets is the total probability of the particle to be detected anywhere after the detection attempt, which is just one from the normalization of the wave function.
\section{Non-local initial conditions} \label{appendix:nonlocal}
Throughout the paper, we've assumed that $\ket{\psi_{in}}$ is a localized state for the sake of simplicity, however it should be noted that our method also works for a non-local initial wave functions after some slight modifications which will be detailed in this appendix.
The first and most important of these modifications is that rather than simply setting $\ket{\rho(0)} = \ket{\psi_{in}}$ we'll first need to localize the wave function using the measurement at time $\tau$.
This step is needed since the stochastic matrix $G$ can only be used to evolve the system in time if the wave function is localized.
\begin{equation}
    \ket{\rho(\tau)} = \sum_{x \in X} |\braket{x|e^{-iH\tau}|\psi_{in}}|^2\ket{x}
    .
\end{equation}
After finding the probability vector at time $\tau$ we'll subtract the first detection probability $F_1=\braket{\psi_{tar}|\rho(\tau)}$ from it:
\begin{equation}
    \ket{\rho(\tau^+)} = \sum_{x \in X/\{\psi_{tar}\}} |\braket{x|e^{-iH\tau}|\psi_{in}}|^2 \ket{x}
    .
\end{equation}

Once this process is done the wave function is localized to some site $x_1$ and we can simply continue to evolve this probability vector in time using $G$ as if it were the initial state of our system.
However, it should be noted that if we simply calculate the generating function from this probability vector as it is, the indices of the probabilities will be shifted.
\begin{equation}
    \widetilde{F(z)}' = \sum_{n=1}^{\infty} F_{n+1} z^n = \frac{\braket{\psi_{tar}|\widetilde{G(z)}|\rho(\tau^+)}}{1 + \braket{\psi_{tar}|\widetilde{G(z)}|\psi_{tar}}}
    .
\end{equation}
We can compensate for this by multiplying the generating function by $z$ and adding $z F_1$ to it.
\begin{equation}
    \widetilde{F(z)} = z F_1 + z\widetilde{F(z)}'
    .
\end{equation}
Once this step is done we can use the generating function in just the same manner as we would in the case where $\ket{\psi_{in}}$ is localized.
However, it should also be noted that the general results we've derived in Sec. \ref{sec:momentresults} no longer apply.

\section{Discontinues jumps of $\braket{n}$ in the transition problem \label{appendix:n^1 disjump}}
In this appendix we derive a general formula for the value of $\braket{n}$ at an exceptional sampling rate when it doesn't diverge and demonstrate that when not diverging it often discontinuously jumps.
We'll start by examining the behavior of the function $g(z)$ defined in Eq. (\ref{generalBKN}) in the limit where $z \to 1$ and $\tau \to \tau_{\mbox{\footnotesize{ex}}}$ where $\tau_{\mbox{\footnotesize{ex}}}$ is an exceptional sampling rate of the system.
As a reminder, the function $g(z)$ is:
\begin{equation}
\begin{split}
    g(z)
    =&\sum_{\lambda} \sum_{k=1}^{g_{\lambda}} \sum_{j=1}^{g_1} \frac{
    \lambda z
    }{
    1-\lambda z
    }
    f(\ket{\lambda_{k}}, \ket{1_{j}})
\\
    f(\ket{\lambda_{k}}, \ket{1_{j}})
    =&\braket{\psi_{tar}|1_{j}} \braket{1_{j}|\psi_{in}} |\braket{\psi_{tar}|\lambda_{k}}|^2\\
    -&\braket{\psi_{tar}|\lambda_{k}} \braket{\lambda_{k}|\psi_{in}} |\braket{\psi_{tar}|1_{j}}|^2
    .
\end{split}
\end{equation}

The problematic terms in the sum over $\lambda_{k}$ are the terms whose eigenvalue is one.
If the inner sum over $j$ is non-zero for any of them then $\braket{n}$ will diverge.
Assuming that it's zero and then simplifying we get the following equation for the value of $\braket{n}$ at an exceptional point which doesn't cause it to diverge:
\begin{equation}
    \label{exepBKN}
    \small
    \braket{n}_{\mbox{\footnotesize{ex}}} = A^{-1} + A^{-2}
    \sum_{\lambda \neq 1} \sum_{k=1}^{g_{\lambda}} \sum_{j=1}^{g_1}  \frac{\lambda}{1-\lambda} f(\ket{\lambda_{k}}, \ket{1_{j}})
    .
\end{equation}
Where $A=\sum_{k=1}^{g_1} |\braket{\psi_{tar}|1_k}|^2$.
Although the assumption that the inner sum is zero may seem unlikely, it is satisfied fairly often.
We show an example of this in Fig. \ref{fig:HexagonTransitionAverage}, this discontinues jump is fairly common in other systems as well.

One additional assumption we need to make is that for at least one of the eigenstates whose eigenvalue is one besides $\ket{\phi}$ its projection onto the target state is non-zero:  $\braket{\psi_{tar}|\lambda_k} \neq 0$. Without this assumption we'll get $\braket{n}_{\mbox{\footnotesize{ex}}} = \braket{n}$.
We think that this assumption is justified by the fact that the projection of the target site onto the eigenstates of $G$ appears in both the denominator and enumerator of the generating function, so if the projections of all of these eigenstates are zero this means that this set of eigenstates had no influence on the statistics of the first detection time to begin with.
It is fairly simple to invent a Hamiltonian describing a graph where some parts of the graph cannot be reached from others.
For any such Hamiltonian  it is clear that the eigenstates describing one part of the system will have no effect on the behaviour of a walk on a different part which cannot be reached, meaning that all the projections would be zero.
In such cases our assumption is wrong and the exceptional sampling rate in question will have no effect on the statistics of the measurement induced quantum walk.
Keep in mind that all that was shown in this section does not mean that $\braket{n}$ necessarily discontinuously jumps if it doesn't diverge, as the two expressions can still equal each other.
However given the assumption that the projection onto the set of exceptional eigenstates is non-zero, there is also no particular reason why the two expressions for $\braket{n}$ should equal each other, so more often than not they will be different as can be seen in Fig. \ref{fig:HexagonTransitionAverage}

\section{Two level system}
\label{appendix:twolevelsystem}
In this appendix we solve the first detection problem for a 2 level system in detail, so as to present the steps one would needs to perform to use our formalism in practice.
The Hamiltonian of the two level system is given by:
\begin{equation}
\label{eq:2lvlSystemHamiltonian}
    H = -\gamma (\ket{0} \bra{1} + \ket{1} \bra{0}) + U \ket{1} \bra{1}
    .
\end{equation}
As previously mentioned in Sec. \ref{section:model}, this Hamiltonian can describe a particle hopping between two distinct sites or any arbitrary two state quantum system where one state has a higher energy than the other, such as a spin 1/2 particle in a magnetic field where the measurement is of the orientation of the spin ($X = \{ \ket{L} \text{, } \ket{R} \}$) for example.
Note that our method is only applicable to this example if the axis of measurement is not parallel to the orientation of the magnetic field, since in that case the states being measured are the eigenstates of the Hamiltonian, which would cause $G$ to simply be the identity matrix.

The first step to finding either the return or transition probabilities is to diagonalize $G$. To do this we first diagonalize the Hamiltonian to obtain the time evolution of every initial condition ($\ket{0}$ and $\ket{1}$) and then we use those results in Eq. (\ref{eq:Gdefinition}) to compute $G$:
\begin{equation}
\small{
    G=
    \begin{pmatrix}
    \frac{U^2+2\gamma^2(1 + \cos(\tau \sqrt{U^2+4\gamma^2}))}{U^2+4\gamma^2} & \frac{2\gamma^2(1 - \cos(\tau \sqrt{U^2+4\gamma^2}))}{U^2+4\gamma^2}
    \\
    \frac{2\gamma^2(1 - \cos(\tau \sqrt{U^2+4\gamma^2}))}{U^2+4\gamma^2} & \frac{U^2+2\gamma^2(1 + \cos(\tau \sqrt{U^2+4\gamma^2}))}{U^2+4\gamma^2}
    \end{pmatrix}
}
    .
\end{equation}
After this step is done we diagonalize $G$ to find it's eigenstates and eigenvalues.
\begin{equation}
\label{2latticeEigenvalues}
\begin{split}
        \ket{\lambda_1} &= \frac{1}{\sqrt{2}} (\ket{0} + \ket{1}) = \ket{\phi}
        \\
        \ket{\lambda_2} &= \frac{1}{\sqrt{2}} (-\ket{0} + \ket{1})
        \\
        \lambda_1 &= 1
        \\
        \lambda_2 &= \frac{U^2+4\gamma^2 \cos(\tau \sqrt{U^2+4\gamma^2})}{U^2+4\gamma^2}
\end{split}
\end{equation}
Notice that $|\lambda_2| \leq 1$ as expected. Exceptional sampling rates are found when the Cosine found in $\lambda_2$ equals one which causes the eigenvalue itself to equal one.

Using these we can now easily calculate the first detection generating function Eq. (\ref{eq:ogGenFuncDefinition}) for any transition on the system in the $\ket{0} \text{, } \ket{1}$ basis using Eq. (\ref{GeneratingFunction}).
Note that since the on site energy $U$ only appears squared both possible transition problems on this graph behave identically, same for the two possible return problems.
Using Eq. (\ref{2latticeEigenvalues}) and the aforementioned remark that the Cosine should equal one we have
\begin{equation}
\label{eq:2lvlsysExceptionals}
    \frac{4 \pi^2 k^2}{ \tau^2} = U^2 +4\gamma^2
\end{equation}
where $k$ is a non-zero natural number (assuming $\gamma \neq 0$).
Choosing values of $\tau$, $\gamma$, and $U$ which satisfy Eq. (\ref{eq:2lvlsysExceptionals}) causes the transition matrix $G$ to become the identity matrix.
From this it's easy to see that for exceptional sampling rates the particle will be detected at the first attempt with certainty. For non-exceptional sampling rates we can use the general results derived in Sec. \ref{sec:momentresults} which tell us that in the return problem $P_{det} = 1$ and $\braket{n} = 2$.
These results can also be confirmed by evaluating the generating function and its derivative directly.
We can obtain the variance using either Eq. (\ref{eq:varianceformula}) or by taking the second derivative of $\widetilde{F(z)}$.
Either way we'll find that it goes like $\Delta n^2 \sim (1-\lambda_2)^{-1}$. Note that although the variance goes to infinity as $\lambda_2$ goes to 1, when $\lambda_2$ is 1 the variance becomes zero.
This is because at those sampling rates the particle is detected at the first attempt with probability one.

\begin{figure}[h]
  \centering
  \includegraphics[width=0.5\textwidth]{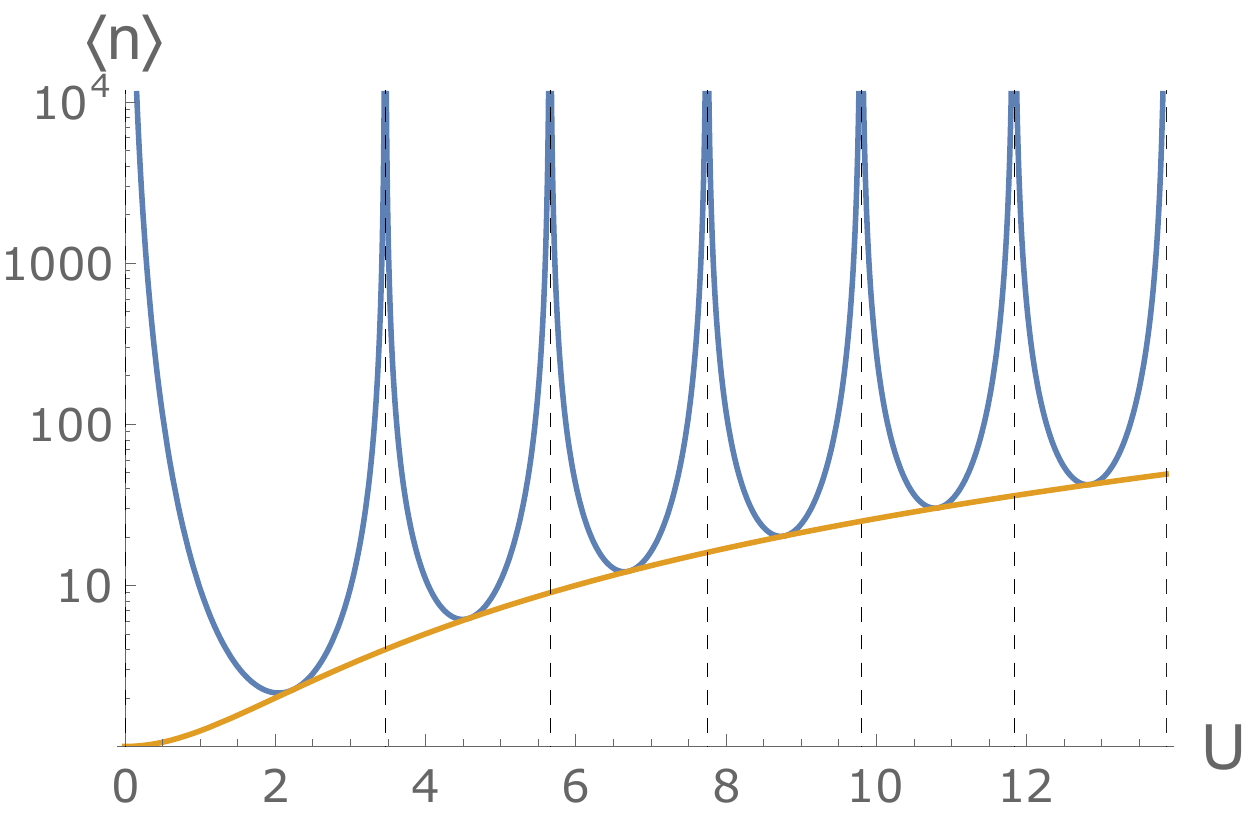}
  \caption{The average number of measurements until first detection in the transition from $\ket{\psi_{in}} = \ket{0}$ to $\ket{\psi_{tar}} = \ket{1}$ for the two level system model whose Hamiltonian is given in Eq. (\ref{eq:2lvlSystemHamiltonian}) as a function of the on site energy $U$ plotted in blue where we set $\tau = \pi$ and $\gamma = 1$. The average diverges at every exceptional value of $U$ while also tending to increase overall like $U^{2}/4$ (plotted in orange). The exceptional energy differences are denoted by the dashed vertical lines. The variance (not plotted) behaves very similarly.}
  \label{fig:2siteAvgN}
\end{figure}

Next, we'll briefly examine the transition from $\ket{0}$ to $\ket{1}$. Keep in mind that since $|\braket{0|e^{-i \tau H}|1}|^2=|\braket{1|e^{-i \tau H}|0}|^2$ the behaviour of this transition is identical to the one form $\ket{1}$ to $\ket{0}$ and the choice to examine one and not the other is arbitrary.
As expected based on Sec. \ref{sec:momentresults} we find that $P_{det}$ is 1 for all non-exceptional combinations of $U$, $\gamma$ and $\tau$ and that for the exceptional ones the particles is never detected.
In addition, we find that the average and variance both diverge near these values, though at the values themselves they jump to zero since all probabilities become zero.
We plot the average as a function of the on site energy is Fig. \ref{fig:2siteAvgN}.

\section{Infinite line return problem probability vector derivation}
\label{app:inflinederivationappendix}
In this appendix we derive the Fourier transform of the probability vector for the return problem on the infinite line lattice whose Hamiltonian is given by Eq. (\ref{eq:inflineH}) and whose initial condition is $\ket{\psi_{in}} = \ket{0}$.
We start with the general form of the probability vector at time $t = \tau n$ which, for the time evolution operator give in Eq. (\ref{eq:inflinetimeevomatrix}), is given by:
\begin{equation}
    \ket{\rho (\tau n)} =\left( \sum_{x, x'=-\infty}^{\infty} |J_{x-x'}(2\gamma \tau)|^2 \ket{x}\bra{x'} \right)^n \ket{0}
    .
\end{equation}
Notice that the expression we've arrived at is the discrete convolution of $|J_{x}(2\gamma \tau)|^2$ with itself $n$ times with respect to $x$.
Using the discrete convolution theorem, its Fourier transform equals the $n$th power of the Fourier transform of $|J_{x}(2\gamma \tau)|^2$.
To find the Fourier transform of $|J_{x}(2\gamma \tau)|^2$ we'll start with the generating function of the Bessel functions:
\begin{equation}
    e^{\frac{z}{2}(t-\frac{1}{t})} = \sum_{x=-\infty}^{\infty}t^x J_x(z)
    .
\end{equation}
Setting $t=e^{i\theta}$ and integrating from zero to $2\pi$ we get:
\begin{equation}
    \label{eq:J0}
    \frac{1}{2\pi}\int_{0}^{2\pi}e^{iz \sin(\theta)}d\theta = J_0(z)
    .
\end{equation}
Next, we make two copies of the generating function.
In one copy we set $t=e^{i \theta}$, and in the other copy we take the complex conjugate and set $t=e^{i ( \theta + k)}$.
We then multiply these two by each other to obtain:
\begin{equation}
    e^{iz( \sin(\theta) - \sin(\theta + k))} = \sum_{x, y=-\infty}^{\infty}e^{-iky} e^{i\theta(x-y)} J_x(z) (J_y(z))^*
    .
\end{equation}
Dividing by $2 \pi$ and integrating from zero to $2\pi$ with respect to $\theta$ we get:
\begin{equation}
    \frac{1}{2\pi}\int_{0}^{2\pi}e^{iz( \sin(\theta) - \sin(\theta + k))}d\theta = \sum_{y=-\infty}^{\infty}e^{-iky} |J_y(z)|^2
    .
\end{equation}
In order to evaluate the integral on the left hand side, we make use of the fact that we can write $\sin(\theta) - \sin(\theta + k)$ as $a \sin(\theta + b)$ where $a = 2 \sin(k / 2)$, we don't need to find $b$ since the integral is over the whole cycle anyways
\begin{equation}
    \label{eq:sinsum}
    \frac{1}{2\pi}\int_{0}^{2\pi}e^{2iz \sin(k/2) \sin(\theta + b)}d\theta = \sum_{y=-\infty}^{\infty}e^{-iky} |J_y(z)|^2
    .
\end{equation}
Using Eqs. (\ref{eq:J0}) and (\ref{eq:sinsum}), we find that the Fourier transform is:
\begin{equation}
    \sum_{y=-\infty}^{\infty}e^{-iky} |J_y(2 \gamma \tau)|^2 = J_0(4 \gamma \tau \sin(\frac{k}{2}))
    .
\end{equation}
Raising this expression to the $n$th power, we find that the Fourier transform of the probability vector at time $t = \tau n$ is:
\begin{equation}
    \sum_k e^{-ikx} \braket{x|\rho(\tau n)} = J_0(4 \gamma \tau \sin(\frac{k}{2}))^n
\end{equation}
Taking the inverse Fourier transform, this shows that the probability vector is given by:
\begin{equation}
    \braket{x|\rho(\tau n)} = \frac{1}{2 \pi} \int_0^{2\pi} e^{ikx} J_0(4 \gamma \tau \sin(\frac{k}{2}))^n
\end{equation}

\section{Saddle point approximations derivation}
\label{appendixSaddlepoint}
In this appendix we derive the results presented in Sec. \ref{secSadlePoint}.
These solutions are based on the methods presented in \cite{10.1214/aoms/1177728652}, and can be briefly summarized in that we can obtain an approximate expression for the probability vector using:
\begin{equation}
\label{eqMainSaddlePointEquation}
\begin{split}
    P_{x,n} &\approx \frac{1}{\sqrt{2 \pi K''(\hat{u})}} \exp{(K(\hat{u}) -\hat{u} x)}
    \\
    K(u) &= \ln{\braket{\exp{(ux)}}} = n \ln{(I_0 (4 \gamma \tau \sinh{(u/2)}))}
\end{split}
\end{equation}
Where $\hat{u}$ satisfies $K'(\hat{u}) = x$.
In our case it is the solution to the equation:
\begin{equation}
\label{eqKDerivativeOGEquation}
    x = 2 n \gamma \tau \frac{I_1 (4 \gamma \tau \sinh{(\hat{u}/2)})}{I_0 (4 \gamma \tau \sinh{(\hat{u}/2)})} \cosh{(\hat{u}/2)}
\end{equation}
Looking at the region where $x \sim n$ by setting $x = 2 n \gamma \tau l$ we have the equation:
\begin{equation}
\label{EqMainThingToApproxInHere}
    l = \frac{I_1 (4 \gamma \tau \sinh{(\hat{u}/2)})}{I_0 (4 \gamma \tau \sinh{(\hat{u}/2)})} \cosh{(\hat{u}/2)}
\end{equation}
We'll solve this equation for different regions of the probability vector by using different appropriate approximations.
Where neither approximation applies, we solve Eq. (\ref{eqKDerivativeOGEquation}) numerically.

\subsection{Small $l$ solution}
\label{appendixsubsectionsmoll}
This solution corresponds to the center of the distribution, where from both the central limit theorem and the previously presented Edgeworth series solution we expect to find a Gaussian distribution.
For a small $l$ we can approximate the right hand side of Eq. (\ref{EqMainThingToApproxInHere}) using a first order Taylor series, which gives us:
\begin{equation}
\begin{split}
    l &= \gamma \tau \hat{u}
    \\
    \hat{u} &= \frac{x}{2n \gamma^2 \tau^2} = \frac{x}{\Delta x^2}
\end{split}
\end{equation}
As a reminder, $x$ and $\Delta x^2$ are unitless.
Plugging this result into Eq. (\ref{eqMainSaddlePointEquation}) and adding one additional minor approximation, we obtain a Gaussian distribution as expected:
\begin{equation}
\begin{split}
    P_{x,n} &= \frac{1}{\sqrt{2 \pi \Delta x^2}} \exp (-\frac{x^2}{2 \Delta x^2})
    \\
    P_{x,n} &= \frac{1}{\sqrt{4 \pi n \gamma^2 \tau^2}} \exp (-\frac{x^2}{4 n \gamma^2 \tau^2})
\end{split}
\end{equation}

\subsection{Large $l$ solution}
\label{appendixsubsectionbigel}
This solution corresponds to the far tails of the distribution, we expect it to display very strong decay to zero.
For this region the right hand side of Eq. (\ref{EqMainThingToApproxInHere}) converges to just the hyperbolic cosine and we get:
\begin{equation}
\begin{split}
    l &= \cosh (\hat{u}/2)
    \\
    \hat{u} &= 2 \text{ arccosh}(l)
\end{split}
\end{equation}
Plugging these into Eq. (\ref{eqMainSaddlePointEquation}), we make use of the following additional approximations:
\begin{equation}
    \frac{I_1(4 \gamma \tau l)}{I_0(4 \gamma \tau l)} \approx \frac{I_2(4 \gamma \tau l)}{I_0(4 \gamma \tau l)} \approx 1
\end{equation}
This gives us the following approximation for the probability vector:
\begin{equation}
\label{EqBigZProbVec}
    P_{x,n} = \frac{1}{
    2 \pi n \gamma \tau \sqrt{\left( \frac{x}{2 \gamma \tau n} \right)^2 - 1}
    }\frac{
    I_0 \left( 4 \gamma \tau \sqrt{\left( \frac{x}{2 \gamma \tau n} \right)^2 - 1} \right)^n
    }{
    \exp \left( 2 x \text{ arccosh} \left( \frac{x}{2 \gamma \tau n} \right) \right)
    }
\end{equation}
In the limit we took of $1 \ll l$, we by extension also have $1 \ll l \ll n \ll x$, meaning that the term is the denominator in Eq. (\ref{EqBigZProbVec}) is much greater than the one in the numerator, demonstrating the rapid decay to zero that we expected.

\begin{widetext} 
\section{Derivation of the $\braket{n}$ and $\Delta n^2$ equations}
\label{app:derivationLONG}
In this appendix we show the derivation of Eq. (\ref{eq:avgnfinalform}, \ref{eq:varianceformula}) in detail, as well as a more general formula for $\Delta n^2$ than what we presented in Sec. \ref{sec:momentresults}.
As a reminder, we are starting with Eq. (\ref{eq:expandedgenfunc}) in the $\ket{\psi_{in}} = \ket{\psi_{tar}}$ case and our goal is to find the limit as $z \to 1$ of its first and second derivative with respect to $z$.
For the purposes of this section, we introduce some new notation in order to shorten the following equations.
First, we abbreviate the sums over the eigenstates $\sum_{\lambda} \sum_{k=1}^{g_{\lambda}} \ket{\lambda_k}$ as just $\sum_{\lambda} \ket{\lambda}$.
Each such sum is over all eigenstates unless stated otherwise, I.E. $\sum_{\lambda = 1}$ is only over eigenstates whose eigenvalue is one and $\sum_{\lambda \neq 1}$ is over all other eigenvalues.
Secondly, we abbreviate the squared projection of the initial state $\ket{\psi_{in}}$ onto each eigenstate as $\psi_{\lambda} = |\braket{\psi_{in}|\lambda}|^2$.

Starting with $\braket{n}$, we we add and subtract one to the numerator in Eq. (\ref{eq:expandedgenfunc}) and rewrite it using the shorthand notation we've introduced:
\begin{equation}
    \widetilde{F(z)} = \frac{
    1 + \sum_{\lambda} \psi_{\lambda} \frac{\lambda z}{1-\lambda z} - 1
    }{
    1 + \sum_{\lambda} \psi_{\lambda} \frac{\lambda z}{1-\lambda z}
    } = 1 - \frac{1}{
    1 + \sum_{\lambda} \psi_{\lambda} \frac{\lambda z}{1-\lambda z}
    }
    .
\end{equation}
Taking the first derivative, we get:
\begin{equation}
\label{eq:genfuncfirstder}
    \frac{d}{dz} \widetilde{F(z)} = \frac{
    \sum_{\lambda} \psi_{\lambda} \frac{\lambda}{(1-\lambda z)^2}
    }{
    (1 + \sum_{\lambda} \psi_{\lambda} \frac{\lambda z}{1-\lambda z})^2
    }
    .
\end{equation}
Next, we multiply and divide the generating function by $(1-z)^2$
\begin{equation}
    \frac{d}{dz} \widetilde{F(z)} = \frac{
    \sum_{\lambda} \psi_{\lambda} \lambda \frac{(1-z)^2}{(1-\lambda z)^2}
    }{
    (1-z + \sum_{\lambda} \psi_{\lambda} \lambda z \frac{1-z}{1-\lambda z})^2
    }
    .
\end{equation}
Written like this, we can see that the reason exceptional eigenvalues are significant is that only the eigenstates whose eigenvalue is one will remain after we take the limit $z \to 1$.
Taking the limit, we easily obtain Eq. (\ref{eq:avgnfinalform}).
\begin{equation}
\label{eq:appendixeqavgfinal}
    \lim_{z \to 1} \frac{d}{dz} \widetilde{F(z)} = \frac{
    \sum_{\lambda = 1} \psi_{\lambda}
    }{
    (\sum_{\lambda = 1} \psi_{\lambda})^2
    } = \frac{1}{\sum_{\lambda = 1} \psi_{\lambda}}
    .
\end{equation}

For the variance, we start by first relating it to $\braket{n}$ and the second derivative of the generating function.
\begin{equation}
\label{eq:variancebasicstructure}
    \Delta n^2 = \braket{n^2} - \braket{n}^2
    = \frac{d}{dz} (z \frac{d}{dz}) \widetilde{F(z)}|_{z \to 1} - \braket{n}^2
    = \frac{d^2}{dz^2} \widetilde{F(z)}|_{z \to 1} +\braket{n} - \braket{n}^2
    .
\end{equation}
Next, we need to find the value of the second derivative at $z \to 1$.
Taking the derivative of Eq. (\ref{eq:genfuncfirstder}) we get:
\begin{equation}
\label{eq:genfuncsecondder}
\begin{split}
    \frac{d^2}{dz^2} \widetilde{F(z)} = &\Bigg{[}
    \sum_{\lambda} \psi_{\lambda}
    \frac{2 \lambda^2}{(1-\lambda z)^3}
    +
    \sum_{\lambda, \Lambda} \psi_{\lambda} \psi_{\Lambda}
    \frac{2 \lambda^2}{(1-\lambda z)^3} \frac{\Lambda z}{1 - \Lambda z}
    \\
    &-
    2 \sum_{\lambda, \Lambda} \psi_{\lambda} \psi_{\Lambda}
    \frac{\lambda}{(1-\lambda z)^2} \frac{\Lambda}{(1-\Lambda z)^2}
    \Bigg{]}
    \Big{(} 1 + \sum_{\lambda} \psi_{\lambda} \frac{\lambda z}{1-\lambda z} \Big{)}^{-3}
    .
\end{split}
\end{equation}
Since the limit of a sum is the sum of the limits if both limits exist, we can simplify our equation slightly by first finding the limit of the first sum which appears in the numerator and later find the limit of the rest of the sums.
We can compute this first limit easily in the same way we did for Eq. (\ref{eq:appendixeqavgfinal}) by multiplying and dividing by $(1-z)^3$ so that after taking the limit only the eigenstates whose eigenvalue is one will remain.
\begin{equation}
\label{eq:genfuncsecondder2}
    \lim_{z \to 1} \frac{
    \sum_{\lambda} \psi_{\lambda}
    \frac{2 \lambda^2}{(1-\lambda z)^3}
    }{
    (1 + \sum_{\lambda} \psi_{\lambda} \frac{\lambda z}{1-\lambda z})^3
    }
    =
    \lim_{z \to 1} \frac{
    \sum_{\lambda} \psi_{\lambda}
    2 \lambda^2 \frac{(1-z)^3}{(1-\lambda z)^3}
    }{
    (1-z + \sum_{\lambda} \psi_{\lambda} \lambda z \frac{1-z}{1-\lambda z})^3
    }
    =
    \frac{
    2
    }{
    (\sum_{\lambda = 1} \psi_{\lambda})^2
    }
    =
    2 \braket{n}^2
    .
\end{equation}
Next, to find the limit of the two double sums we'll first combine them into a single double sum.
As we do this we also multiply the whole expression by the $(1-z)^3$ which originates in the denominator of Eq. (\ref{eq:genfuncsecondder})
\begin{equation}
\label{eq:genfuncsecondder3}
\begin{split}
    (1-z)^3
    \Bigg{[}
    &\sum_{\lambda, \Lambda} \psi_{\lambda} \psi_{\Lambda}
    \frac{2 \lambda^2}{(1-\lambda z)^3} \frac{\Lambda z}{1 - \Lambda z}
    -
    2 \sum_{\lambda, \Lambda} \psi_{\lambda} \psi_{\Lambda}
    \frac{\lambda}{(1-\lambda z)^2} \frac{\Lambda}{(1-\Lambda z)^2}
    \Bigg{]}
    \\
    =
    & \sum_{\lambda, \Lambda} \psi_{\lambda} \psi_{\Lambda}
    \frac{\lambda \Lambda}{(1-\lambda z)^3} \frac{(1-z)^3}{(1-\Lambda z)^2}
    \left(
    2 \lambda z (1-\Lambda z) - 2 (1 - \lambda z)
    \right)
    \\
    =
    & \sum_{\lambda, \Lambda} \psi_{\lambda} \psi_{\Lambda}
    \frac{\lambda \Lambda}{(1-\lambda z)^3} \frac{(1-z)^3}{(1-\Lambda z)^2}
    \left(
    4 \lambda z - 2 \lambda \Lambda z^2 - 2
    \right)
    .
\end{split}
\end{equation}
In order to see what elements of this double sum reduce to zero in the limit and which will remain, we'll break it apart into 4 sums depending on whether the eigenvalues of the eigenstates being summed over equal one or not.
\begin{equation}
\label{eq:genfuncsecondder4}
\begin{split}
    & \sum_{\substack{\lambda = 1 \\ \Lambda = 1}} \psi_{\lambda} \psi_{\Lambda}
    \frac{1}{(1-z)^3} \frac{(1-z)^3}{(1-z)^2}
    \left(
    4 z - 2 z^2 - 2
    \right)
    \\ +
    & \sum_{\substack{\lambda = 1 \\ \Lambda \neq 1}} \psi_{\lambda} \psi_{\Lambda}
    \frac{\Lambda}{(1- z)^3} \frac{(1-z)^3}{(1-\Lambda z)^2}
    \left(
    4 z - 2 \Lambda z^2 - 2
    \right)
    \\ +
    & \sum_{\substack{\lambda \neq 1 \\ \Lambda = 1}} \psi_{\lambda} \psi_{\Lambda}
    \frac{\lambda}{(1-\lambda z)^3} \frac{(1-z)^3}{(1- z)^2}
    \left(
    4 \lambda z - 2 \lambda z^2 - 2
    \right)
    \\ +
    & \sum_{\substack{\lambda \neq 1 \\ \Lambda \neq 1}} \psi_{\lambda} \psi_{\Lambda}
    \frac{\lambda \Lambda}{(1-\lambda z)^3} \frac{(1-z)^3}{(1-\Lambda z)^2}
    \left(
    4 \lambda z - 2 \lambda \Lambda z^2 - 2
    \right)
    .
\end{split}
\end{equation}
We can see here that in the limit only the first two sums will remain whereas the other two will go to zero.
After taking the limit and simplifying we get:
\begin{equation}
\label{eq:genfuncsecondder5}
    -2 \sum_{\substack{\lambda = 1 \\ \Lambda = 1}} \psi_{\lambda} \psi_{\Lambda} +
    2 \sum_{\substack{\lambda = 1 \\ \Lambda \neq 1}} \psi_{\lambda} \psi_{\Lambda}
    \frac{\Lambda}{1-\Lambda}
    .
\end{equation}
Putting this result together with Eq. (\ref{eq:genfuncsecondder2}) in Eq. (\ref{eq:variancebasicstructure}) we obtain a general formula for the variance which is correct for all sampling rates:
\begin{equation}
    \Delta n^2 =
    \braket{n} +
    \braket{n}^2 +
    \braket{n}^3
    (
    -2\sum_{\substack{\lambda = 1 \\ \Lambda = 1}} \psi_{\lambda} \psi_{\Lambda} +
    2 \sum_{\substack{\lambda = 1 \\ \Lambda \neq 1}} \psi_{\lambda} \psi_{\Lambda}
    \frac{\Lambda}{1-\Lambda}
    )
    .
\end{equation}
For non exceptional sampling rates, the only term in the sum over eigenstates whose eigenvalue is one is the $\ket{\phi}$ term which is just $|X|^{-1}$ and the expression can be simplified to Eq. (\ref{eq:varianceformula}).
\end{widetext}
\bibliography{references.bib}% Produces the bibliography via BibTeX.
\end{document}